\def\Box{\kern1pt\vbox{\hrule height 1.2pt\hbox{\vrule width 1.2pt\hskip 3pt
   \vbox{\vskip 6pt}\hskip 3pt\vrule width 0.6pt}\hrule height 0.6pt}\kern1pt}
\def\gtwid{\mathrel{\raise.3ex\hbox{$>$\kern-.75em\lower1ex\hbox{$\sim$}}}}
\def\ltwid{\mathrel{\raise.3ex\hbox{$<$\kern-.75em\lower1ex\hbox{$\sim$}}}}
\def\Box{\kern1pt\vbox{\hrule height 1.2pt\hbox{\vrule width 1.2pt\hskip 3pt
   \vbox{\vskip 6pt}\hskip 3pt\vrule width 0.6pt}\hrule height 0.6pt}\kern1pt}
\documentstyle[epsfig,12pt]{article}

\begin{document}
\begin{titlepage}
\begin{flushright}
astro-ph/9811430 \\ UFIFT-HEP-98-18
\end{flushright}
\vspace{.4cm}
\begin{center}
\textbf{One Loop Back Reaction On Chaotic Inflation}
\end{center}
\begin{center}
L. R. Abramo$^{* \dagger}$ and R. P. Woodard$^{\ddagger}$
\end{center}
\begin{center}
\textit{Department of Physics \\ University of Florida \\ 
Gainesville, FL 32611 USA}
\end{center}
\begin{center}
ABSTRACT
\end{center}
\hspace*{.5cm} We extend, for the case of a general scalar potential, the 
inflaton-graviton Feynman rules recently developed by Iliopoulos {\it et al.} 
\cite{Ilio}. As an application we compute the leading term, for late co-moving
times, of the one loop back reaction on the expansion rate for $V(\varphi) = 
\frac12 m^2 \varphi^2$. This is expressed as the logarithmic time derivative
of the scale factor in the coordinate system for which the expectation value
of the metric has the form: $\langle 0 \vert g_{\mu\nu}({\overline t},{\vec x})
\vert 0 \rangle dx^{\mu} dx^{\nu} = - d{\overline t}^2 + a^2({\overline t}) 
d{\vec x} \cdot d{\vec x}$. This quantity should be a gauge independent 
observable. Our result for it agrees exactly with that inferred from the effect
previously computed by Mukhanov {\it et al.} \cite{Mukh,Abra} using canonical 
quantization. It is significant that the two calculations were made with 
completely different schemes for fixing the gauge, and that our computation was
done using the standard formalism of covariant quantization. This should settle
some of the issues recently raised by Unruh \cite{Unru}.
\begin{flushleft}
PACS numbers: 04.60.-m, 98.80.Cq
\end{flushleft}
\vspace{.4cm}
\begin{flushleft}
$^*$ Address after May 1, 1999: Theoretische Physik, Ludwig Maximilians \\
Universit\"{a}t, Theresienstr. 37, D-80333 M\"{U}NCHEN, GERMANY \\
$^{\dagger}$ e-mail: abramo@phys.ufl.edu \\
$^{\ddagger}$ e-mail: woodard@phys.ufl.edu
\end{flushleft}
\end{titlepage}

\section{Introduction}

We wish to address a controversy which has arisen in the literature of 
scalar-driven inflation. The dispute concerns the recent claim by Mukhanov,
A\-bram\-o and Brandenberger \cite{Mukh,Abra} that infrared modes can generate 
a significant one loop back-reaction which reduces the expansion rate over 
the course of inflation. Unruh \cite{Unru} has raised a number of serious 
questions about their methodology and the plausibility of their conclusion.

We begin by summarizing Unruh's objections:
\begin{enumerate}
\item It is difficult to understand how long wavelength modes can affect the
local geometry since they should appear spatially constant to a local observer.
\item To leading order in the long wavelength expansion the mode solutions are
all equivalent, locally, to coordinate transformations which can have no effect
on local invariants.
\item The quantization procedure employed by Mukhanov {\it et al.} is suspect
because their dynamical variable is nonzero for only one of the two leading
long wavelength solutions. Since this dynamical variable possesses another,
independent solution, the corresponding degree of freedom must be unphysical.
\item What Mukhanov {\it et al.} refer to as ``gauge independent'' quantities
are really just the local dynamical variables in a particular gauge.
\item Mukhanov {\it et al.} employ an unconventional variation of perturbation
theory in which the effective stress-energy tensor of the first order equations
renormalizes the zeroth order stress-energy tensor.
\item The contributions to the metric at second order --- the zero mode of 
which is what Mukhanov {\it et al.} computed --- depend upon the gauge chosen 
for expressing the first order terms.
\end{enumerate}

Without wishing to criticize good people who addressed an important issue at
the extreme limit of their formalism's applicability, we must admit to a
certain sympathy for Unruh's methodological objections. One of the present 
authors also had difficulty understanding the work of Mukhanov {\it et al.} on 
account of (4-6).\footnote{Other confusing points were the characterization of 
what is obviously a one loop effect as ``classical'' and the attribution of 
this effect to an instability in the classical energy functional (which is 
actually stable).} However, we believe the physics of what they did is correct,
and that is the point of this paper. 

After fixing notation about the perturbative background in Section 2 we 
comment in Section 3 on the physics of the process and we partially address 
Unruh's objections (1-3). The remainder of the paper is devoted to checking 
the calculation of Mukhanov {\it et al.}, in a completely different gauge, 
using the standard formalism of covariant quantum field theory. The Feynman 
rules are given in Section 4. These were lifted from a recent paper by 
Iliopoulos, Tomaras, Tsamis and Woodard \cite{Ilio}, which we have extended 
so that the propagators can be computed (as mode sums) for a general scalar 
potential. Section 5 attaches the external lines (which are retarded 
propagators in Schwinger's formalism \cite{Schw}) needed to convert the 
amputated 1-point functions into the expectation values of the metric and 
the scalar. We also explain how these expectation values are used to compute
physical observables which measure the cosmological expansion rate and the
evolution of the scalar. In Section 6 we give the procedure used for isolating
the leading contribution to each propagator from superadiabatically amplified
modes at late co-moving times. This is the chief physical approximation of the 
paper. The amputated 1-point functions are computed in Section 7 and processed
to give the two physical observables. Section 8 summarizes the various results.

\section{The perturbative background}

The system under study is that of general relativity with a general,
minimally coupled scalar:
\begin{equation}
{\cal L} = {1 \over 16\pi G} R \sqrt {-g} - \frac12 \partial_{\mu} 
\varphi \partial_{\nu} \varphi g^{\mu \nu} \sqrt {-g} - V(\varphi) 
\sqrt {-g} \; . \label{eq:action}
\end{equation}
This section concerns the homogeneous and isotropic backgrounds $g_0$ and 
$\varphi_0$ about which perturbation theory will be formulated. Three 
classes of identities turn out to be interesting for our purposes:
\begin{enumerate}
\item Those which are exact and valid for any potential $V(\varphi)$;
\item Those which are valid in the slow roll approximation but still for 
any potential; and 
\item Those which are valid for the slow roll approximation with the 
potential $V(\varphi) = \frac12 m^2 \varphi^2$.
\end{enumerate}
We shall develop them in this order, identifying the point at which each
further specialization and approximation is made.

Among the exact identities is the relation between co-moving and conformal
coordinates:
\begin{equation}
ds_0^2 = -dt^2 + a_0^2(t) d{\vec x} \cdot d{\vec x} = \Omega^2(\eta) 
\left\{-d\eta^2 + d{\vec x} \cdot d{\vec x}\right\} \; .
\end{equation}
This implies:
\begin{equation}
dt = \Omega d\eta \qquad , \qquad a_0(t) = \Omega(\eta) \; .
\end{equation}
The Hubble ``constant'' is the logarithmic co-moving time derivative of the 
background scale factor:
\begin{equation}
H \equiv {\dot{a}_0 \over a_0} = {\Omega^{\prime} \over \Omega^2} \; ,
\end{equation}
where a dot denotes differentiation with respect to (background) co-moving
time and a prime stands for differentiation with respect to conformal time.

Two of Einstein's equations are nontrivial in this background:
\begin{eqnarray}
3 H^2 & = & \frac12 \kappa^2 \left\{\frac12 \dot{\varphi}_0^2 + V(\varphi_0)
\right\} \; ,\\
-2 \dot{H} - 3 H^2 & = & \frac12 \kappa^2 \left\{\frac12 \dot{\varphi}_0^2
- V(\varphi_0)\right\} \; ,
\end{eqnarray}
where $\kappa^2 \equiv 16 \pi G$ is the loop counting parameter of perturbative
quantum gravity. One can use the two Einstein equations to derive the scalar
equation of motion:
\begin{equation}
\ddot{\varphi}_0 + 3 H \dot{\varphi}_0 + V_{,\varphi}(\varphi_0) = 0 \; ,
\end{equation}
where $V_{,\varphi} \equiv {\partial V}/{\partial \varphi}$. One can also
invert the Einstein equations to solve for the Hubble constant and its first
derivative:
\begin{eqnarray}
H^2 & = & \frac16 \kappa^2 \left\{\frac12 \dot{\varphi}_0^2 + V(\varphi_0)
\right\} \; , \\
\dot{H} & = & -\frac14 \kappa^2 \dot{\varphi_0}^2 \; .
\end{eqnarray}

Sometimes it is more convenient to write the scalar quantities in terms of 
the Hubble constant and its derivative:
\begin{eqnarray}
\dot{\varphi}_0^2 & = & -{4 \over \kappa^2} \dot{H} \; ,\\
V(\varphi_0) & = & {2 \over \kappa^2} (\dot{H} + 3 H^2) \; .
\end{eqnarray}
At other times one wants to express the scalar quantities using the conformal 
factor $\Omega$:
\begin{eqnarray}
\dot{\varphi}_0^2 & = & {4 \over \kappa^2} {1 \over \Omega^2} \left\{-
{\Omega^{\prime\prime} \over \Omega} + 2 \left({\Omega^{\prime} \over 
\Omega}\right)^2\right\} \; ,\\
V(\varphi_0) & = & {2 \over \kappa^2} {1 \over \Omega^2} \left\{
{\Omega^{\prime\prime} \over \Omega} + \left({\Omega^{\prime} \over 
\Omega}\right)^2\right\} \; .
\end{eqnarray}
And the conformal time derivative of the scalar $\varphi_0^{\prime} = \Omega
\dot{\varphi}_0$ is also useful:
\begin{equation}
{\varphi_0^{\prime}}^2 = {4 \over \kappa^2} \left\{-{\Omega^{\prime\prime}
\over \Omega} + 2 \left({\Omega^{\prime} \over \Omega}\right)^2 \right\}\; .
\end{equation}

Successful models of inflation require the following two conditions which 
define the {\it slow roll approximation}:
\begin{eqnarray}
\vert \ddot{\varphi}_0 \vert & \ll & H \vert \dot{\varphi}_0 \vert \; ,
\label{eq:slow1} \\
\dot{\varphi}_0^2 & \ll & V(\varphi_0) \; . \label{eq:slow2}
\end{eqnarray}
It follows that there are two small parameters. Although these are 
traditionally expressed as ratios of the potential and its derivatives the more
useful quantities for our work are ratios of the Hubble constant and its 
derivatives:
\begin{equation}
{-\dot{H} \over H^2} \ll 1 \qquad , \qquad {\vert \ddot{H} \vert \over - H 
\dot{H}} \ll 1 \; .
\end{equation}
For models of interest to us the rightmost of these parameters is negligible 
with respect to the leftmost one. We shall also assume that the derivative of
the scalar is negative:
\begin{equation}
\varphi_0^{\prime} = -{2 \over \kappa} \Omega \sqrt{-\dot{H}} \; .
\end{equation}

The slow roll approximation gives useful expansions for simple calculus
operations. For example, ratios of derivatives of the field are:
\begin{eqnarray}
{\varphi_0^{\prime\prime}\over \varphi_0^{\prime}} & = & H \Omega \left( 1 +
\frac12 {\ddot{H} \over H \dot{H}} \right) \; ,\\
{\varphi_0^{\prime\prime\prime}\over \varphi_0^{\prime}} & = & 2 H^2 \Omega^2
\left(1 + {\dot{H} \over 2 H^2} + \dots\right) \; .
\end{eqnarray}
Successive partial integration also defines useful slow roll expansions:
\begin{eqnarray}
\int dt H^{\alpha} \Omega^{\beta} & = & {1 \over \beta} H^{\alpha-1} 
\Omega^{\beta} \left\{1 + {(\alpha - 1) \over \beta} \left({-\dot{H} \over H^2}
\right) + \dots\right\} \; ,\\
\int dt H^{\alpha} & = & {1 \over \alpha + 1} {H^{\alpha+1} \over \dot{H}}
\left\{1 + {1 \over \alpha + 2} {H \ddot{H} \over \dot{H}^2}+ \dots\right\} \; .
\end{eqnarray}

In discussing the physical significance of their result Mukhanov, Abramo and 
Brandenberger specialized to the simplest potential for chaotic inflation
\cite{Lind}:
\begin{equation}
V(\varphi) \rightarrow \frac12 m^2 \varphi^2 \; . \label{eq:partV}
\end{equation}
In the slow roll approximation with this potential one can solve explicitly 
for the scalar's evolution:
\begin{equation}
\varphi_0(t) = \varphi_i - {2 \over \sqrt{3}} {m \over \kappa} t \; .
\end{equation}
The interesting geometrical quantities have the following expressions in
terms of $\varphi_0(t)$:
\begin{eqnarray}
\dot{H} & \approx & -\frac13 m^2 \; ,\\
H & \approx & {1 \over \sqrt{12}} \kappa m \varphi_0(t) \; .
\end{eqnarray}
Note that $\ddot{H} \approx 0$ for this potential, so only one of the slow roll
parameters is nonzero. The slow roll approximation implies that the initial 
value of the scalar field is much larger than the Planck mass:
\begin{equation}
\varphi_i \gg {1 \over \kappa} \; .
\end{equation}
Inflation ends in this model when $\varphi_0(t) \sim \kappa^{-1}$.

\section{Physical comments}

The physical mechanism behind what Mukhanov, Abramo and Brandenberger 
\cite{Mukh,Abra} have found for scalar-driven inflation is roughly the same as 
that studied previously by Tsamis and Woodard \cite{Tsam} in the context of 
inflation caused by a bare cosmological constant. There is such a simple 
physical model for what is going on that we would be derelict in our duty of 
explication not to present it. Formalists should rest assured that this is 
merely a qualitative description of phenomena whose reality has already been 
established by computing what should be invariant observables in the standard 
formalism of covariant quantization.

Owing to the rapid expansion of spacetime and the special properties of the 
dynamical quanta involved,\footnote{These properties are (1) effective
masslessness on the Hubble scale and (2) the absence of classical conformal 
invariance.} there is a vast enhancement of the 0-point energy which the
uncertainty principle requires to be present in each dynamical degree of
freedom. This is the phenomenon of {\it superadiabatic amplification}, first 
studied by Grishchuk \cite{Gris}. A simple picture for it is that virtual 
pairs with wavelengths comparable to the horizon can become trapped in the
expansion of spacetime and not be able to recombine.

Superadiabatic amplification is not a large effect by itself. Although the
total energy contained in infrared modes increases quite rapidly, the 
corresponding expansion of the 3-volume keeps the energy {\it density}
constant for pure de Sitter expansion.\footnote{For certain models of 
scalar-driven inflation the infrared energy density can grow as the scalar
rolls. This is what seems to distinguish those scalar potentials for which 
there is a one loop effect from those for which there is not. The two loop 
effect of pure gravity --- and presumably also gravity with scalars --- does 
not depend upon such growth.} For this background it is simple to show that
there is only about one extra infrared quantum per Hubble volume. The 
interesting, secular effect derives from the gravitational interaction between 
these quanta. As each virtual pair is pulled apart, its gravitational 
potentials fill the intervening space. These remain to add with those of the 
next pair. Even though the 0-point energy stays constant, the induced 
gravitational potential increases. It is the interaction energy between this 
and the 0-point energy, and between the gravitational potentials themselves,
which gives the effect. 

In the purely gravitational model of Tsamis and Woodard, linearized gravitons 
can only induce gravitational potentials at second order in the weak field 
expansion. Since superadiabatic amplification is a one loop effect this means 
that the secular back-reaction comes at two loop order. When inflation is 
driven by a scalar field its quanta can induce gravitational potentials even 
at linearized order in the weak field expansion. This is why Mukhanov, Abramo 
and Brandenberger were able to follow what is essentially the same physical 
process with a vastly simpler one loop calculation.

Either way, the effect is to slow inflation because gravity is attractive.
Since gravity is also a weak interaction, even for GUT scale inflation, the 
process requires an enormous amount of time before it can become significant. 
A direct consequence is that the equation of state of the induced stress 
tensor must be approximately that of negative vacuum energy. To see this 
consider the relation implied by conservation between the induced energy 
density $\rho(t)$ and the induced pressure $p(t)$:
\begin{equation}
\dot{\rho} = -3 H \left(\rho + p\right) \; .
\end{equation}
Since the accumulation of a significant effect requires many Hubble times, 
$\vert \dot{\rho} \vert \ll H \vert \rho\vert$, and it must be that $p(t)$ 
nearly cancels $\rho(t)$.

A sometimes confusing point is that one does not require the complete theory
of quantum gravity in order to study an infrared process such as this. As
long as spurious time dependence is not injected through the ultraviolet
regularization, the late time back-reaction is dominated by ultraviolet 
finite, nonlocal terms whose form is entirely controlled by the low energy
limiting theory. This theory must be general relativity, with the possible 
addition of some light scalars. It is worth commenting that infrared phenomena 
can always be studied using the low energy effective theory. This is why Bloch 
and Nordsieck \cite{BlNo} were able to resolve the infrared problem of QED 
before the theory's renormalizability was suspected. It is also why Weinberg 
\cite{Wein} was able to achieve a similar resolution for quantum general 
relativity with zero cosmological constant. And it is why Feinberg and Sucher 
\cite{FeSu} were able to compute the long range force due to neutrino exchange 
using Fermi theory. More recently Donoghue \cite{Dono} has been working along 
the same lines for quantum gravity with zero cosmological constant.

We emphasize that the process is causal, in spite of its close association 
with modes whose wavelengths have redshifted beyond the horizon. This emerges 
most clearly in the two loop computation of Tsamis and Woodard where the 
effect derives from integrating interaction vertices over the past lightcone 
of the point at which the expansion rate is being measured. Because gravitons 
are massless these interactions superpose coherently. Because gravitons are 
not conformally invariant they reflect the enormous physical volume of the 
past lightcone rather than its minuscule conformal volume. The growth in the 
back-reaction is directly attributable to the fact that the invariant volume 
of the past lightcone increases without bound as one observes at later and 
later times. 

Causality is also built into the work of Mukhanov, Abramo and Brandenberger
through their use of the Heisenberg field operators. The equations of motion
for these are simply operator realizations of the causal field equations of 
classical general relativity. In a local gauge one can express an operator at 
the spacetime point $(t,{\vec x})$ entirely in terms of the operators and their 
time derivatives on that part of the initial value surface which lies on or 
within the past lightcone of $(t,{\vec x})$. There might be some dispute about 
what this means nonperturbatively, where the quantum metric can have a 
significant impact on the lightcone, but it makes perfect sense in the 
perturbative regime under study. The time dependence Mukhanov, Abramo and 
Brandenberger obtain derives in part from the continual redshift of new modes 
from ultraviolet to infrared but mostly from the growth of the infrared mode 
functions which can occur in some (but not all) inflationary backgrounds.

We turn now to issues (1-3) listed in our Introduction. Regarding the first 
objection, it is relevant to note that while long wavelength modes indeed 
appear spatially constant to a local observer, so too does the cosmological 
expansion rate. Therefore the causative agent and its purported effect are 
commensurate. 

Viewed from the perspective of obtaining a long period of inflation it is 
rather local, short wavelength phenomena that ought to be regarded with
suspicion. Without severe fine tuning the natural duration of any process
mediated by short wavelength quanta must be the Hubble time or less. It seems
reasonable to conclude that a mechanism for screening the cosmological 
constant must also end inflation if nothing else does the job first. But 
inflation has to persist for many Hubble times in order to explain the large
scale smoothness of the observed universe. Note that infrared, long wavelength 
phenomena can require much longer to produce a significant effect because they
can act by coherently superposing an inherently weak interaction over the past 
lightcone. It is only an enormous expansion of the invariant volume contained
in the past lightcone that can compensate for the weakness of gravitational 
self-interactions.

Finally, one must distinguish between a local observation of the cosmological 
expansion rate, and the local expansion rate that would be produced by spatial 
inhomogeneities in the vacuum energy. Many people believe that whatever is 
suppressing the former must also suppress the latter. We do not share this 
view. Experiment is sadly unable to decide the matter but it seems to us that 
a local fluctuation which created a large enough region of negative $\rho + 3 
p$ should result in that region beginning to undergo inflation. We believe 
that known physical principles already suffice to explain why such 
fluctuations are rare in the observed universe \cite{Vach}. Were it otherwise 
one would not be able to make conventional models of inflation agree with 
observation by the unaesthetic device of fine tuning the bare cosmological 
constant which is, be it noted, spatially homogeneous.

The relevant point about the second objection is that the ``stress tensor'' 
of gravitational perturbation theory is not an invariant, or even a scalar. 
So the fact that infrared mode solutions look, to leading order for small wave 
number, like coordinate transformations does not mean they necessarily have no 
effect. Superadiabatic amplification allows these modes to carry nonzero
energy and pressure in spite of their extreme redshift. The proper way to 
determine their effect is by computing the metric's response at second order 
and then addressing the gauge issue of what this response means physically.
We believe this is what Mukhanov, Abramo and Brandenberger did, although 
perhaps not as transparently as one might wish. To check their result we made 
what ought to be the same computation in a completely different gauge and using
the standard formalism of covariant quantization, and we got the same answer.

This is the right point to comment on the gauge issue, which was also raised 
extensively by Unruh. What both we and Mukhanov {\it et al.} computed was the
expectation value of the metric in the presence of a particular state and
in a fixed gauge. There is no doubt that this quantity depends upon the gauge
in which the computation was done. It is important to realize that a quantity
is not automatically devoid of physical import by virtue of being gauge 
dependent. It can still contain useful physical information which can be
separated from the unphysical, gauge dependent part. Examples of this abound
in quantum field theory. The most straightforward is the way in which gauge 
dependent Green's functions can be processed, using LSZ reduction, to give
gauge independent, on-shell scattering amplitudes. (This is discussed in any
standard text on quantum field theory, for example that by Peskin and 
Schroeder \cite{QFT}.) One does not even have to consider products of field
operators. There is an elegant formalism, due to DeWitt \cite{DeWi}, in which 
the S-matrix is obtained from the in-out matrix element of the dynamical 
variable in the presence of a general scattering state.

So the expectation value of the metric contains valid physical information;
the question is how to extract it. Our technique exploits the special property 
of the initial state of being homogeneous and isotropic. This means that a 
co-moving coordinate system exists for which:
\begin{equation}
\langle 0 \vert g_{\mu \nu}({\overline t},{\vec x}) dx^{\mu} dx^{\nu} \vert 0 
\rangle = -d{\overline t}^2 + a^2({\overline t}) d{\vec x} \cdot d{\vec x} \; .
\end{equation}
Our observable is the logarithmic time derivative of the scale factor in this
coordinate system:
\begin{equation}
H_{\rm eff}({\overline t}) \equiv {1 \over a({\overline t})} {d a({\overline 
t}) \over d{\overline t}} \; .
\end{equation}
One can investigate how this quantity changes under a variation of the gauge 
fixing functional and the result is that it does not change \cite{Tsam}. This 
would seem to be the analog of DeWitt's theorem about the gauge independence 
of the on-shell S-matrix. Of course the absence of gauge dependence does not 
automatically endow a quantity with physical import. We interpret $H_{\rm 
eff}({\overline t})$ as the expansion rate a local observer would measure in 
the presence of state $\vert 0 \rangle$. It certainly has this meaning in the 
classical limit but we are willing to entertain dissident views.

We come finally to Unruh's doubts about the formalism of Mukhanov, Feldman and 
Brandenberger \cite{MFB}. His argument is based on the long wavelength 
solution \cite {Unru} he found for the linearized Newtonian potential, which 
we shall call $n(x)$. In our notation this quantity corresponds to the 
following invariant element:
\begin{equation}
g_{\mu\nu} dx^{\mu} dx^{\nu} = \Omega^2 \left\{- (1 + 2 \Omega^{-1} n) 
d\eta^2 + (1 - 2 \Omega^{-1} n) d{\vec x} \cdot d{\vec x}\right\} \; .
\end{equation}
When the linearized $g_{0i}$ equations are used to eliminate the scalar field
the linearized $g_{00}$ equation becomes:
\begin{equation}
n^{\prime\prime} - \nabla^2 n - 2 {\varphi_0^{\prime\prime} \over \varphi_0^{
\prime}} n^{\prime} + \left({\Omega^{\prime\prime} \over \Omega} - 2 
{{\Omega^{\prime}}^2 \over \Omega^2}\right) n = 0 \; .
\end{equation}
In the limit that the $\nabla^2$ term can be neglected Unruh obtains the
following independent solutions:
\begin{eqnarray}
n_1 & = & \Omega - {\Omega^{\prime} \over \Omega^2} \int_{-\infty}^{\eta} 
d{\overline \eta} \; \Omega^2({\overline \eta}) \; , \\
n_2 & = & {\Omega^{\prime} \over \Omega^2} \; .
\end{eqnarray}

Unruh's problem concerns what happens when $n_1$ and $n_2$ are substituted
into the dynamical variable used by Muk\-han\-ov, Feldman and Brandenberger:
\begin{equation}
v = {4 \over \kappa^2} {1 \over \varphi_0^{\prime}} \left\{n^{\prime} +
\left(-{\Omega^{\prime\prime} \over \Omega^{\prime}} + 2 {\Omega^{\prime} \over
\Omega}\right) n\right\} \; . \label{eq:vMFB}
\end{equation}
Although $n_1$ produces a reasonable function:
\begin{equation}
v_1 = {4 \over \kappa^2} {1 \over \varphi_0^{\prime}} \left\{-{\Omega \Omega^{
\prime\prime} \over \Omega^{\prime}} + 2 \Omega^{\prime}\right) = {\Omega^2
\over \Omega^{\prime}} \varphi_0^{\prime} \; ,
\end{equation}
the $n_2$ solution gives $v = 0$! This is disturbing because the formalism of
Mukhanov, Feldman and Brandenberger quantizes $v$ as a scalar field which obeys
the following second order equation:
\begin{equation}
v^{\prime\prime} - \nabla^2 v - {v_1^{\prime\prime} \over v_1} v = 0 \; .
\label{eq:veqn}
\end{equation}
In the limit that the $\nabla^2$ term can be neglected one finds that $v = v_1$
is indeed a solution, as is:
\begin{equation}
v_2(\eta) = v_1(\eta) \int_{\eta}^0 {d{\overline \eta} \over v^2_1({
\overline \eta})} \; .
\end{equation}
Since this second solution does not correspond to any combination of Unruh's
two long wavelength solutions he concludes that it must be unphysical and
that the formalism is therefore suspect.

In fact neither the $v_2$ solution nor the $n_2$ solution is unphysical, they
simply correspond to different orders in the long wavelength expansion.
At fixed, nonzero wave number ${\vec k}$ one can express the two independent 
solutions for the Newtonian potential as power series in $k^2$. The zeroth 
terms in these two series are Unruh's solutions, $n_1(\eta)$ and $n_2(\eta)$. 
However, one should really include some higher order terms as well since the 
physical relevance of the solutions is for modes with small but nonzero wave 
number.\footnote{There are a lot more of these. The number of zero modes is 
constant in time whereas the inflationary redshift eventually makes the 
physical wave number ($\Omega^{-1} {\vec k}$) of {\it any} mode small. Further,
as Unruh pointed out, the ${\vec k} = 0$ system is degenerate in that his two 
solutions become unphysical on account of being exactly coordinate 
transformations. The one physical solution is paradoxically absent from the 
${\vec k} \neq 0$ system. It appears because the $g_{0i}$ equation is 
automatically satisfied for ${\vec k} = 0$ and therefore fails to relate the 
zero modes of the Newtonian potential and the scalar \cite{Unru}.} The first 
order correction to $n_2$ can be expressed using an advanced Green's function:
\begin{equation}
N_2(\eta,k) = n_2(\eta) +{4 k^2 \over \kappa^2} \int_{\eta}^0 d{\overline 
\eta} \left\{n_1(\eta) n_2({\overline \eta}) - n_2(\eta) n_1({\overline \eta})
\right\} {n_2({\overline \eta}) \over \left(\varphi_0^{\prime}({\overline 
\eta})\right)^2} + O(k^4) \; .
\end{equation}
When it is substituted for $n(x)$ in (\ref{eq:vMFB}) the result is:
\begin{equation}
v\vert_{n = N_2} = {4 k^2 \over \kappa^2} v_2(\eta) + O(k^4) \; .
\end{equation}
So the solution sets of the two variables are in one-to-one correspondence and
there is no obvious problem with the formalism of Mukhanov, Feldman and 
Brandenberger.

At the price of specializing to power law inflation one can even see how the 
$n$ and $v$ solution sets relate to all orders. For $\Omega = (\eta_0/\eta)^{
s/(s-1)}$ the two mode solutions for the Newtonian potential are proportional 
to Bessel functions of order $\mu = \frac12 + 1/(s-1)$. Unruh's solutions are
the zeroth order terms in the power series expansions of the following:
\begin{eqnarray}
N_1(\eta,k) & = & {\Gamma(1-\mu) \over s + 1} \left({k \eta_0 \over 2}
\right)^{\mu} \sqrt{{\eta_0 \over \eta}} J_{-\mu}(k \eta) \; , \\
N_2(\eta,k) & = & -{s \over s-1} \Gamma(1+\mu) \left({k \eta_0 \over 2}
\right)^{-\mu} {1 \over \sqrt{\eta_0 \eta}} J_{\mu}(k \eta) \; .
\end{eqnarray}
One can easily verify that the solutions to (\ref{eq:veqn}) which make
contact with $v_1$ and $v_2$ are:
\begin{eqnarray}
V_1(\eta,k) & = & {2 \over \kappa} {\Gamma(-\mu) \over \sqrt{s}} \left({k 
\eta_0 \over 2}\right)^{\mu+1} \sqrt{\eta \over \eta_0} J_{-\mu-1}(k \eta) \; ,
\\
V_2(\eta,k) & = & -{\kappa \over 4} \sqrt{s} \Gamma(1+\mu) \left({k \eta_0 
\over 2} \right)^{-\mu-1} \sqrt{\eta_0 \eta} J_{\mu+1}(k \eta) \; .
\end{eqnarray}
For power law inflation the relation (\ref{eq:vMFB}) between $n$ and $v$ is
recognizable as the recursion relation which produces $-J_{\mu+1}$ from $J_{
\mu}$ (and $+J_{-\mu-1}$ from $J_{-\mu}$):
\begin{equation}
v \rightarrow -{2 \over \kappa} {s -1 \over \sqrt{s}} \sqrt{\eta} \left\{
[\sqrt{\eta} n]^{\prime} - {\mu \over \eta} [\sqrt{\eta} n]\right\} \; ,
\nonumber 
\end{equation}
and the $V_i$'s descend from the $N_i$'s as follows:
\begin{eqnarray}
V_1(\eta,k) & = & -{2 \over \kappa} {s -1 \over \sqrt{s}} \sqrt{\eta} 
\left\{[\sqrt{\eta} N_1(\eta,k)]^{\prime} - {\mu \over \eta} [\sqrt{\eta} 
N_1(\eta,k)]\right\} \; , \\
V_2(\eta,k) & = & {\kappa^2 \over 4 k^2} \times -{2 \over \kappa} {s -1 \over 
\sqrt{s}} \sqrt{\eta} \left\{[\sqrt{\eta} N_2(\eta,k)]^{\prime} - {\mu \over
\eta} [\sqrt{\eta} N_2(\eta,k)]\right\} \; .
\end{eqnarray}

\section{Feynman rules}

The purpose of this section is to give the Feynman rules for the general
inflaton-graviton action (\ref{eq:action}). We have mostly borrowed these
from a recent paper by Iliopoulos, Tomaras, Tsamis and Woodard \cite{Ilio}. 
The one exception concerns the issue of mixing between the scalar and the 
$00$ component of the graviton field. Iliopoulos {\it et al.} were only 
able to solve the system for a class of backgrounds including those of power 
law inflation, but not the power law backgrounds typical of chaotic inflation. 
We have achieved a general solution. One should also note that all formulae 
given in this section are exact. We have made neither the slow roll 
approximation nor have we specialized to the case of a quadratic potential. Of 
course that will be necessary in order to convert the formal mode sums into 
explicit results, but the task of making these approximations has been 
postponed to the end of Section 5.

Our quantum fields are the scalar $\phi$ and the conformally rescaled
pseudo-graviton $\psi_{\mu\nu}$:
\begin{eqnarray}
\varphi & \equiv & \varphi_0 + \phi \; ,\\
g_{\mu \nu} & \equiv & \Omega^2 \left(\eta_{\mu\nu} + \kappa \psi_{\mu\nu} 
\right) \equiv \Omega^2 {\widetilde g}_{\mu\nu} \; .
\end{eqnarray}
It should be noted that cosmologists typically restrict the word ``graviton'' 
to that part of the metric which interpolates dynamical spin two quanta at 
linearized order. Adhering to this convention would be terrifically cumbersome
in the context of BRS quantization beyond linearized order. Our 
``pseudo-graviton'' also includes degrees of freedom which are constrained or 
pure gauge. This is the standard usage in particle theory, cf. the ``photon'' 
field, propagator and interactions of QED and the ``gluon'' field, propagator 
and interactions of QCD \cite{QFT}. We shall try to avoid misunderstandings, 
without over-burdening the notation, by following the convention of Iliopoulos 
{\it et al.} who parenthesized the word ``general'' before ``pseudo-graviton.''

As usual, (general) pseudo-graviton indices are raised and lowered with the 
Lorentz metric. After many tedious partial integrations the invariant 
Lagrangian can be written as a total derivative plus the following:
\begin{eqnarray}
\lefteqn{{\cal L}_{\rm inv} = \sqrt {- \widetilde g} {\widetilde g}^{
\alpha\beta} {\widetilde g}^{\rho\sigma} {\widetilde g}^{\mu\nu} } 
\nonumber \\
& & \mbox{} \times \left\{ \frac12 \psi_{\alpha \rho , \mu} \psi_{\nu 
\sigma , \beta}  -\frac12 \psi_{\alpha \beta, \rho} \psi_{\sigma \mu , \nu} 
+ \frac14 \psi_{\alpha \beta , \rho} \psi_{\mu\nu ,\sigma} - \frac14 
\psi_{\alpha\rho ,\mu} \psi_{\beta\sigma ,\nu} \right\} \Omega^2 
\nonumber \\
& & \mbox{} - \frac12 \sqrt {- \widetilde g} {\widetilde g}^{\rho\sigma} 
{\widetilde g}^{\mu\nu} \psi_{\rho\sigma ,\mu} \psi_{\nu}^{~\alpha} 
(\Omega^2)_{,\alpha}  - \Omega^2 \varphi_0^{\prime} \phi_{,\mu} {\widetilde 
g}^{0\mu} \sqrt{-{\widetilde g}} \nonumber \\
& & \mbox{} - \frac12 \Omega^2 \phi_{,\mu} \phi_{,\nu} {\widetilde g}^{
\mu\nu} \sqrt{-{\widetilde g}} - \sum_{n=1}^{\infty} {1 \over n!} 
{\partial^n V(\varphi_0) \over \partial \varphi^n} \Omega^4 \phi^n 
\sqrt{-{\widetilde g}} \; , \label{eq:invariant}
\end{eqnarray}
where a comma denotes differentiation. Gauge fixing is accomplished by
adding a gauge fixing functional and the corresponding ghost action to
obtain the BRS Lagrangian:
\begin{equation}
{\cal L}_{\rm BRS} = {\cal L}_{\rm inv} - \frac12 \eta^{\mu \nu} F_{\mu}
F_{\nu} - \Omega \; {\overline \omega}^{\mu} {\delta F}_{\mu} \; .
\end{equation}
The symbol ${\delta F}_{\mu}$ represents the variation of the gauge fixing
functional under an infinitesimal diffeomorphism parameterized by the ghost
field $\omega_{\mu}$. We will follow Iliopoulos {\it et al.} in our choice
of gauge fixing functional:
\begin{equation}
F_{\mu} = \Omega \left(\psi_{\mu~ ,\nu}^{~\nu} - \frac12 \psi_{,\mu} - 2 
{\Omega^{\prime} \over \Omega} \psi_{\mu 0} + \eta_{\mu 0} \kappa 
{\varphi}_0^{\prime} \phi \right) \; .
\end{equation}
A great advantage of this gauge is that it decouples the tensor structure of
the propagators from their dependence on spacetime. The propagator becomes
a small number of constant tensors multiplying only three different types of
mode sums. Another advantage is that the limit $\Omega \rightarrow 1$ takes
this gauge to one of the standard gauges of flat space, which often provides a
useful correspondence check.

With a few more partial integrations the terms quadratic in the various 
quantum fields can be reduced to the following form:
\begin{eqnarray}
\lefteqn{{\cal L}_{\rm BRS}^{(2)} = \frac12 \psi^{\mu\nu} D_{\mu\nu}^{~~~\rho
\sigma} \psi_{\rho\sigma} + \psi^{\mu\nu} \Omega \left[-\kappa \varphi_0^{
\prime\prime} t_{\mu} t_{\nu}\right] \Omega \phi} \nonumber \\
& & \mbox{} + \frac12 \phi \Omega \left[\partial^2 + \frac14 \kappa^2 
{\varphi_0^{\prime}}^2 + {\varphi_0^{\prime\prime\prime} \over 
\varphi_0^{\prime}} \right] \Omega \phi + {\overline \omega}^{\mu} 
\left[{\overline \delta}_{\mu}^{~\nu} {\rm D}_A - t_{\mu} t^{\nu} 
{\rm D}_B\right] \omega_{\nu} \; .
\end{eqnarray}
A number of pieces of notation require explanation. The differential
operators ${\rm D}_A$ and ${\rm D}_B$ are:
\begin{eqnarray}
{\rm D}_A & \equiv & \Omega \left[\partial^2 + {\Omega^{\prime\prime} \over
\Omega}\right] \Omega \; ,\\
{\rm D}_B & \equiv & \Omega \left[\partial^2 + {(\Omega^{-1})^{\prime\prime} 
\over \Omega^{-1}}\right] \Omega \; ,
\end{eqnarray}
where $\partial^2 \equiv \eta^{\mu\nu} \partial_{\mu} \partial_{\nu}$ is the
d'Alembertian in conformal coordinates. It is worth commenting that ${\rm 
D}_A$ is the kinetic operator for a massless, minimally coupled scalar. The 
kinetic operator for the (general) pseudo-graviton is:
\begin{equation}
D_{\mu \nu}^{~~\rho \sigma} \equiv \left[\frac12 {\overline \delta}_{\mu}^{~(
\rho} {\overline \delta}_{\nu}^{~\sigma)} - \frac14 \eta_{\mu \nu} \eta^{\rho 
\sigma} - \frac12 t_{\mu} t_{\nu} t^{\rho} t^{\sigma}\right] {\rm D}_A -
t_{(\mu} {\overline \delta}_{\nu)}^{~~(\rho} t^{\sigma)} {\rm D}_B + t_{\mu}
t_{\nu} t^{\rho} t^{\sigma} {\rm D}_B \; . \label{eq:kinetic}
\end{equation}
Parenthesized indices are symmetrized, the symbol $t_{\mu}$ denotes:
\begin{equation}
t_{\mu} \equiv \eta_{\mu 0} \qquad , \qquad t^{\mu} = \delta^{\mu}_{~0} \; ,
\end{equation}
and a bar above a Lorentz metric or a Kronecker delta symbol means that the 
zero component is projected out:
\begin{equation}
{\overline \eta}_{\mu \nu} \equiv \eta_{\mu \nu} + t_{\mu} t_{\nu} \qquad ,
\qquad {\overline \delta}_{\mu}^{~\nu} \equiv \delta_{\mu}^{~\nu} + t_{\mu}
t^{\nu} \; .
\end{equation}

The quadratic Lagrangian involves two sorts of mixing: that between the 
spatial trace and $\psi_{00}$ and that between $\psi_{00}$ and the scalar 
$\phi$. The first can be removed by the following simple field redefinition:
\begin{equation}
\psi_{ij} \equiv \zeta_{ij} + \delta_{ij} \zeta_{00} \qquad \; \qquad
\psi_{0i} \equiv \zeta_{0i} \qquad \; \qquad \psi_{00} \equiv \zeta_{00}
\; ,
\end{equation}
where small Latin letters denote spatial indices. In these variables the
quadratic part of the Lagrangian becomes:
\begin{eqnarray}
{\cal L}_{\rm BRS}^{(2)} & = &
\frac12 \zeta_{00} {\rm D}_B \zeta_{00} - \kappa \Omega^2 \varphi_0^{
\prime\prime} \zeta_{00} \phi + \frac12 \phi \Omega \left[\partial^2 + 
\frac14 \kappa^2 {\varphi_0^{\prime}}^2 + {\varphi_0^{\prime\prime\prime}
\over \varphi_0^{\prime}}\right] \Omega \phi \nonumber \\
& & \mbox{} + \frac12 \zeta_{ij} \left(\frac12 \delta_{i (k} \delta_{\ell)j} 
- \frac14 \delta_{ij} \delta_{k\ell}\right) {\rm D}_A \zeta_{k\ell} - 
\frac12 \zeta_{0i} \delta_{ij} {\rm D}_B \zeta_{0j} \nonumber \\
& & \hskip 2cm \mbox{} + {\overline \omega}^{\mu} \left[{\overline 
\delta}_{\mu}^{~\nu} {\rm D}_A - t_{\mu} t^{\nu} {\rm D}_B\right] 
\omega_{\nu} \; . \label{eq:quadratic}
\end{eqnarray}
For a general $a_0(t)$ there is no local change of variables which removes
the mixing between $\zeta_{00}$ and $\phi$ off shell. However, it is easy to 
diagonalize the linearized {\it field equations} which determine the on shell
mode solutions. Canonical quantization of the linearized theory can then be
invoked to expand the original quantum fields in terms of creation and 
annihilation operators. It is straightforward to use these expansions to 
express the propagators as mode sums. At the cost of some mixed propagators we 
will eventually obtain a complete expression of the Feynman rules in terms of 
the field variables $\psi_{\mu\nu}$, ${\overline \omega}^{\mu}$, $\omega_{\nu}$
and $\phi$.

It is simplest to absorb a factor of $\Omega$ into $\zeta_{00}$ and $\phi$:
\begin{equation}
z \equiv \Omega \zeta_{00} \qquad , \qquad f \equiv \Omega \phi \; .
\end{equation}
The linearized equations for the $z$--$f$ system are:
\begin{eqnarray}
\left(\partial^2 + \frac14 \kappa^2 {\varphi_0^{\prime}}^2\right) z -
\kappa \varphi_0^{\prime\prime} f & = & 0 \; , \label{eq:first} \\
-\kappa \varphi_0^{\prime\prime} z + \left(\partial^2 + \frac14 \kappa^2 
{\varphi_0^{\prime}}^2 + {\varphi_0^{\prime\prime\prime} \over \varphi_0^{
\prime}}\right) f & = & 0 \; . \label{eq:second}
\end{eqnarray}
Differentiating (\ref{eq:first}) and adding it to $\frac12 \kappa \varphi_0^{
\prime}$ times (\ref{eq:second}) gives the first of our diagonalized field
equations:
\begin{equation}
\left(\partial^2 + \frac14 \kappa^2 {\varphi_0^{\prime}}^2\right) \left[
z^{\prime} + \frac12 \kappa \varphi_0^{\prime} f\right] = 0 \; .
\end{equation}
The second diagonalized field equation comes from differentiating 
(\ref{eq:second}) and adding it to $\frac12 \kappa \varphi_0^{\prime}$ times
(\ref{eq:first}) minus ${\varphi_0^{\prime\prime}}/{\varphi_0^{\prime}}$ times
(\ref{eq:second}):
\begin{equation}
\left(\partial^2 + \frac14 \kappa^2 {\varphi_0^{\prime}}^2 - {\varphi_0^{
\prime\prime\prime} \over \varphi_0^{\prime}} + 2 {{\varphi_0^{\prime\prime}}^2
\over {\varphi_0^{\prime}}^2}\right) \left[\frac12 \kappa \varphi_0^{\prime}
z + f^{\prime} - {\varphi_0^{\prime\prime} \over \varphi_0^{\prime}} f\right] 
= 0 \; .
\end{equation}

The preceding discussion implies that the diagonal variables are:
\begin{eqnarray}
x(\eta,{\vec x}) & \equiv & z^{\prime}(\eta,{\vec x}) + \frac12 \kappa 
\varphi_0^{\prime}(\eta) f(\eta,{\vec x}) \; , \label{eq:xdef} \\
y(\eta,{\vec x}) & \equiv & \frac12 \kappa \varphi_0^{\prime}(\eta) z(\eta,
{\vec x}) + f^{\prime}(\eta,{\vec x}) - {\varphi_0^{\prime\prime}(\eta) \over 
\varphi_0^{\prime}(\eta)} f(\eta,{\vec x}) \; . \label{eq:ydef}
\end{eqnarray}
Since conformal time derivatives appear in the transformation its inverse
cannot be local in time for the off shell fields. However, by using the 
linearized field equations one can obtain the following expressions for the 
conformal time derivatives of $x$ and $y$:
\begin{eqnarray}
x^{\prime} & = & \left(\nabla^2 + \frac14 \kappa^2 {\varphi_0^{\prime}}^2
\right) z + \frac12 \kappa \varphi_0^{\prime} f^{\prime} - \frac12 \kappa
\varphi_0^{\prime\prime} f \; , \label{eq:xprime} \\
y^{\prime} & = & \frac12 \kappa \varphi_0^{\prime} z^{\prime} - \frac12 \kappa
\varphi_0^{\prime\prime} z - {\varphi_0^{\prime\prime} \over \varphi_0^{
\prime}} f^{\prime} + \left(\nabla^2 + \frac14 \kappa^2 {\varphi_0^{\prime}}^2
+ {{\varphi_0^{\prime\prime}}^2 \over {\varphi_0^{\prime}}^2} \right) f \; . 
\label{eq:yprime}
\end{eqnarray}
Eliminating $z^{\prime}$ and $f^{\prime}$ gives the following on shell
inverse transformation:
\begin{eqnarray}
z = {1 \over \nabla^2} \left[x^{\prime} - \frac12 \kappa \varphi_0^{\prime} y
\right] \; , \label{eq:ztrans} \\
f = {1 \over \nabla^2} \left[-\frac12 \kappa \varphi_0^{\prime} x + y^{\prime}
+ {\varphi_0^{\prime\prime} \over \varphi_0^{\prime}} y\right] \; .
\label{eq:ftrans}
\end{eqnarray}
We stress that since the linearized field equations have been used these 
relations apply only to the on shell mode solutions, not to the off shell 
fields.

The mode equations for all the fields --- including $\Omega \zeta_{ij}$,
$\Omega \zeta_{0i}$, $\Omega {\overline \omega}^{\mu}$ and $\Omega \omega_{
\nu}$ --- can be given a simple, unified treatment. There are three types of 
modes which we shall call $A$, $B$ and $C$. They are defined as the plane 
wave solutions annihilated by the following differential operator:
\begin{equation}
{\cal D}_I \equiv \partial^2 + {\theta_I^{\prime\prime} \over \theta_I} 
\longrightarrow - \left\{{d^2 \over d\eta^2} + k^2 - {\theta_I^{\prime\prime}(
\eta) \over \theta_I(\eta)}\right\}  \; ,
\end{equation}
where $k \equiv \Vert {\vec k} \Vert$ and the various $\theta_I(\eta)$'s are:
\begin{equation}
\theta_A \equiv \Omega \qquad , \qquad \theta_B \equiv \Omega^{-1} \qquad , 
\qquad \theta_C \equiv -{2 \over \kappa} {\Omega^{\prime} \over \Omega^2
\varphi_0^{\prime}} = a_0^{-1} {H \over \sqrt{-\dot{H}}} \; .
\end{equation}
Briefly, the spatial polarizations --- $\Omega \zeta_{ij}$, $\Omega {\overline
\omega}^i$ and $\Omega \omega_j$ --- are comprised of $A$ modes, the mixed 
polarizations --- $\Omega \zeta_{0i}$, $\Omega {\overline \omega}^0$ and
$\Omega \omega_0$ --- are made of $B$ modes, as is the diagonal variable $x$, 
and the other diagonal variable $y$ consists of $C$ modes. Because quantization 
was accomplished by adding a gauge fixing term most of the linearized fields
harbor unphysical quanta. Physical gravitons are $A$ modes that reside in 
$\zeta_{ij}$; the physical scalar is a $C$ mode in $y$.

We will return in 
Section 6 to the problem of obtaining useful approximations for the mode 
functions but we proceed, for now, as though they are known. We define 
$Q_I(\eta,k)$ as descending by perturbative iteration (explained in Section 6)
from the pure negative frequency solution for wave number $k \equiv \Vert 
{\vec k} \Vert$. We also assume it has been canonically normalized:
\begin{equation}
Q_I(\eta,k) Q_I^{*\prime}(\eta,k) - Q_I^{\prime}(\eta,k) Q_I^*(\eta,k) 
= i \; .
\end{equation}

From canonically quantizing the quadratic action (\ref{eq:quadratic}) one
finds that the fields $z(\eta,{\vec x})$ and $z^{\prime}(\eta,{\vec x})$ form
a conjugate pair. The same is true for $f(\eta,{\vec x})$ and $f^{\prime}(
\eta,{\vec x})$, so the only nonzero equal-time commutators involving these 
fields are:
\begin{equation}
\left[z(\eta,{\vec x}),z^{\prime}(\eta,{\vec y})\right] = i \delta^3({\vec x}
- {\vec y}) = \left[f(\eta,{\vec x}),f^{\prime}(\eta,{\vec y})\right] \; .
\end{equation}
From their definitions (\ref{eq:xdef}-\ref{eq:ydef}) and the on shell relations
(\ref{eq:xprime}-\ref{eq:yprime}) one can easily check that the only nonzero
equal-time commutators in the $x$--$y$ sector are:
\begin{equation}
\left[x(\eta,{\vec x}),x^{\prime}(\eta,{\vec y})\right] = - i \nabla^2
\delta^3({\vec x} - {\vec y}) = \left[y(\eta,{\vec x}),y^{\prime}(\eta,
{\vec y})\right] \; .
\end{equation}
We can realize these commutation relations with conventionally normalized 
creation and annihilation operators $(X,X^{\dagger}$ and $(Y,Y^{\dagger})$:
\begin{equation}
\left[X({\vec k}),X^{\dagger}({\vec p})\right] = (2 \pi)^3 \delta^3({\vec k} -
{\vec p}) = \left[Y({\vec k}),Y^{\dagger}({\vec p})\right] \; .
\end{equation}
Since ${\cal D}_B x(\eta,{\vec x}) = 0$ we expand $x$ using $B$ modes:
\begin{equation}
x(\eta,{\vec x}) = \int {d^3k \over (2\pi)^3} k e^{i {\vec k} \cdot {\vec x}}
\left\{X({\vec k}) Q_B(\eta,k) + X^{\dagger}({\vec k}) Q_B^*(\eta,k)\right\}
\; .
\end{equation}
Since ${\cal D}_C y(\eta,{\vec x}) = 0$ we expand $y$ using $C$ modes:
\begin{equation}
y(\eta,{\vec x}) = \int {d^3k \over (2\pi)^3} k e^{i {\vec k} \cdot {\vec x}}
\left\{Y({\vec k}) Q_C(\eta,k) + Y^{\dagger}({\vec k}) Q_C^*(\eta,k)\right\}
\; .
\end{equation}
The on shell transformations (\ref{eq:ztrans}-\ref{eq:ftrans}) allow us 
finally to give operator expansions for $z$ and $f$:
\begin{eqnarray}
z(\eta,{\vec x}) & = & \int {d^3k \over (2\pi)^3} {e^{i {\vec k} \cdot 
{\vec x}} \over k} \left\{-X({\vec k}) Q_B^{\prime}(\eta,k) \right.
\nonumber \\
& & \hskip 3cm \left. + Y({\vec k}) \frac12 \kappa \varphi_0^{\prime}(\eta) 
Q_C(\eta,k) + {\rm c.c.}\right\} \; , \\
f(\eta,{\vec x}) & = & \int {d^3k \over (2\pi)^3} {e^{i {\vec k} \cdot 
{\vec x}} \over k} \left\{X({\vec k}) \frac12 \kappa \varphi_0^{\prime}(\eta) 
Q_B(\eta,k) \right. \nonumber \\
& & \hskip 2cm \left. - Y({\vec k}) \left[{d \over d\eta} + 
{\varphi_0^{\prime\prime}(\eta) \over \varphi_0^{\prime}(\eta)}\right] 
Q_C(\eta,k) + {\rm c.c.}\right\} \; .
\end{eqnarray}

The rest is a standard exercise in free field theory. We choose the state 
$\vert 0 \rangle$ to obey:
\begin{equation}
X({\vec k}) \vert 0 \rangle = 0 = Y({\vec k}) \vert 0 \rangle \; . 
\label{eq:vac}
\end{equation}
The various propagators can be most conveniently expressed in terms of the 
mode sum $i \delta_I(x;x')$:\footnote{Note that we have introduced a time
independent, ultraviolet convergence factor of $e^{-\epsilon k}$. This 
corresponds to an exponential mode cutoff on the initial value surface.}
\begin{eqnarray}
i \delta_I(x;x') & = & -{1 \over \nabla^2} \int {d^3k \over (2\pi)^3} 
e^{i {\vec k} \cdot {\Delta \vec x} - \epsilon k} \left\{\theta(\Delta \eta) 
Q_I(\eta,k) Q_I^*(\eta^{\prime},k) \right. \nonumber \\
& & \hskip 4cm \left. + \theta(-\Delta \eta) Q_I^*(\eta,k) Q_I(\eta^{\prime}
,k)\right\} \; , \\
& = & {1 \over 2 \pi^2} \int_0^{\infty} dk {\sin(k {\Delta x}) \over k 
{\Delta x}} e^{-\epsilon k} \left\{\theta(\Delta \eta) Q_I(\eta,k) 
Q_I^*(\eta^{\prime},k) \right. \nonumber \\
& & \hskip 4cm \left. + \theta(-\Delta \eta) Q_I^*(\eta,k) Q_I(\eta^{\prime},
k)\right\} \; , \label{eq:smallprop}
\end{eqnarray}
where we define the following conformal coordinate differences:
\begin{equation}
{\Delta \eta} \equiv \eta - \eta^{\prime} \qquad , \qquad {\Delta \vec x} 
\equiv {\vec x} - {\vec x}^{\prime} \qquad , \qquad {\Delta x} \equiv
\Vert {\vec x} - {\vec x}^{\prime} \Vert \; .
\end{equation}
Expanding the various time-ordered products and exploiting (\ref{eq:vac})
leads to the following expressions for the propagators:
\begin{eqnarray}
\langle 0 \vert T\left\{z(\eta,{\vec x}) z(\eta^{\prime},{\vec x}^{\prime})
\right\} \vert 0 \rangle & = & {\partial \over \partial \eta} {\partial \over 
\partial \eta^{\prime}} i \delta_B(x;x') \nonumber \\
& & \hskip 1cm \mbox{} + \frac14 \kappa^2 \varphi_0^{\prime}(\eta) 
\varphi_0^{\prime}(\eta^{\prime}) i \delta_C(x;x') \; ,\\
\langle 0 \vert T\left\{z(\eta,{\vec x}) f(\eta^{\prime},{\vec x}^{\prime})
\right\} \vert 0 \rangle & = &- {\partial \over \partial \eta} \frac12 \kappa 
\varphi_0^{\prime}(\eta^{\prime}) i \delta_B(x;x') \nonumber \\
& & \hskip 1cm \mbox{} - \frac12 \kappa \varphi_0^{\prime}(\eta) \left[
{\partial \over \partial \eta^{\prime}} + {\varphi_0^{\prime\prime} \over 
\varphi_0^{\prime}} \right] i \delta_C(x;x') \; , \qquad \\
\langle 0 \vert T\left\{f(\eta,{\vec x}) f(\eta^{\prime},{\vec x}^{\prime})
\right\} \vert 0 \rangle & = & \frac14 \kappa^2 \varphi_0^{\prime}(\eta) 
\varphi_0^{\prime}(\eta^{\prime}) i \delta_B(x;x') \nonumber \\
& & \hskip .5cm \mbox{} + \left[{\partial \over \partial \eta}
+ {\varphi_0^{\prime\prime} \over \varphi_0^{\prime}}\right] \left[{\partial 
\over \partial \eta^{\prime}} + {\varphi_0^{\prime\prime} \over \varphi_0^{
\prime}}\right] i \delta_C(x;x') \; . \qquad 
\end{eqnarray}

Now we restore the factors of $\Omega^{-1}$. These can be used to convert the
conformal time derivatives to co-moving time:
\begin{equation}
\Omega^{-1}(\eta) {\partial \over \partial \eta} = {\partial \over \partial t} 
\qquad , \qquad \Omega^{-1} \varphi_0^{\prime} = \dot{\varphi}_0 \; .
\end{equation}
The three $z$--$f$ propagators become:
\begin{eqnarray}
i \Delta_{\alpha}(x;x') & \equiv & \langle 0 \vert T\left\{\psi_{00}(\eta,
{\vec x}) \psi_{00}(\eta^{\prime},{\vec x}^{\prime})\right\} \vert 0 \rangle
\nonumber \\
& = & {\partial \over \partial t} {\partial \over \partial t'}
i \delta_B(x;x') + \frac14 \kappa^2 \dot{\varphi}_0(t) \dot{\varphi}_0(t')
i \delta_C(x;x') \; , \label{eq:alphaprop} \\
i \Delta_{\beta}(x;x') & \equiv & \langle 0 \vert T\left\{\psi_{00}(\eta,
{\vec x}) \phi(\eta^{\prime},{\vec x}^{\prime})\right\} \vert 0 \rangle
\nonumber \\
& & \hskip -2cm = -{\partial \over\partial t} \frac12 \kappa 
\dot{\varphi}_0(t') i \delta_B(x;x') - \frac12 \kappa \dot{\varphi}_0(t) 
\left[{\partial \over \partial t'} + H + {\ddot{\varphi}_0 \over 
\dot{\varphi}_0}\right] i \delta_C(x;x') \; , \label{eq:betaprop} \\
i \Delta_{\gamma}(x;x') & \equiv & \langle 0 \vert T\left\{\phi(\eta,{\vec x}) 
\phi(\eta^{\prime},{\vec x}^{\prime})\right\} \vert 0 \rangle \nonumber \\
& = & \frac14 \kappa^2 \dot{\varphi}_0(t) \dot{\varphi}_0(t') i \delta_B(x;x') 
\nonumber \\
& & \hskip 1cm \mbox{} + \left[{\partial \over \partial t} + H + 
{\ddot{\varphi}_0 \over \dot{\varphi}_0}\right] \left[{\partial \over \partial 
t'} + H+ {\ddot{\varphi}_0 \over \dot{\varphi}_0}\right] i \delta_C(x;x') \; .
\label{eq:gammaprop}
\end{eqnarray}

The unmixed propagators can also be represented as mode sums:
\begin{equation}
i \Delta_I(x;x) \equiv \Omega^{-1}(\eta) \Omega^{-1}(\eta') (-\nabla^2)
i \delta_I(x;x) \; . \label{eq:bigprop}
\end{equation}
The (general) pseudo-graviton propagator is:
\begin{eqnarray}
\lefteqn{\left\langle 0 \vert T \left\{\psi_{\mu\nu}(x) \psi_{\rho\sigma}(x')
\right\} \vert 0 \right\rangle = i \Delta_A(x;x') 2 \left[{\overline \eta}_{
\mu(\rho} {\overline \eta}_{\sigma)\nu} - {\overline \eta}_{\mu\nu} 
{\overline \eta}_{\rho\sigma}\right]} \nonumber \\
& & \mbox{} -4 i \Delta_B(x;x') t_{(\mu} {\overline \eta}_{\nu)(\rho} 
t_{\sigma)} + i \Delta_{\alpha}(x;x') \left[{\overline \eta}_{\mu\nu} +
t_{\mu} t_{\nu}\right] \left[{\overline \eta}_{\rho\sigma} + t_{\rho} 
t_{\sigma}\right] \; .
\end{eqnarray}
The other propagators are:
\begin{eqnarray}
\left\langle 0 \vert T \left\{\psi_{\mu\nu}(x) \phi(x') \right\} \vert 0 
\right\rangle & = & i \Delta_{\beta}(x;x') \left[{\overline \eta}_{\mu\nu}
+ t_{\mu} t_{\nu}\right] \; , \\
\left\langle 0 \vert T \left\{\phi(x) \phi(x') \right\} \vert 0 \right\rangle 
& = & i \Delta_{\gamma}(x;x') \; , \\
\left\langle 0 \vert T \left\{\omega_{\mu}(x) {\overline \omega}_{\nu}(x') 
\right\} \vert 0 \right\rangle & = & i \Delta_A(x;x') {\overline \eta}_{\mu\nu}
- i \Delta_B(x;x') t_{\mu} t_{\nu} \; .
\end{eqnarray}
All interactions between (general) pseudo-gravitons and scalars can be 
obtained by expanding the invariant Lagrangian (\ref{eq:invariant}) using the
following identities:
\begin{eqnarray}
{\widetilde g}^{\mu\nu} & = & \eta^{\mu\nu} - \kappa \psi^{\mu\nu} + \kappa^2
\psi^{\mu\rho} \psi_{\rho}^{~\nu} - \dots \; , \\
\sqrt{-{\widetilde g}} & = & 1 + \frac12 \kappa \psi + \kappa^2 \left(\frac18
\psi^2 - \frac14 \psi^{\rho\sigma} \psi_{\rho\sigma}\right) + \dots \; .
\end{eqnarray}
Interactions involving ${\overline \omega}^{\mu}$ and $\omega_{\nu}$ can be 
read from the ghost Lagrangian, which we have simplified by neglecting some
total derivatives:
\begin{eqnarray}
{\cal L}_{\rm ghost} & = & - \Omega {\overline \omega}^{\mu} {\delta F}_{\mu}
\; ,\\
& & \hskip -2cm = {\overline \omega}^{\mu} \left[{\overline \delta}_{\mu
}^{~\nu} {\rm D}_A - t_{\mu} t^{\nu} {\rm D}_B\right] \omega_{\nu} + \kappa 
\left(\Omega^2 {\overline \omega}^{\mu}\right)_{,\mu} \left\{\psi^{\nu\rho} 
\omega_{\nu,\rho} + \frac12 \psi^{,\nu} \omega_{\nu} - {\Omega^{\prime} \over 
\Omega} \psi \omega_0\right\} \nonumber \\
& & \mbox{} - \kappa \Omega^2 {\overline \omega}^{\mu , \nu} \left\{\psi_{\mu
\rho} \omega^{\rho}_{~,\nu} + \psi_{\nu\rho} \omega^{\rho}_{~,\mu} + \psi_{
\mu \nu , \rho} \omega^{\rho} -2 {\Omega^{\prime} \over \Omega} \psi_{\mu\nu}
\omega_0\right\} \nonumber \\
& & \hskip 2cm \mbox{} + \kappa^2 \Omega^2 \varphi_0^{\prime} {\overline 
\omega}_0 \phi_{,\nu} \omega^{\nu} \; .
\end{eqnarray}

\section{Attaching external lines}

The purpose of this section is to explain how we pass from the amputated
1-point functions which are actually computed to physical observables. We
begin by expressing the effective Hubble constant in terms of the 
(nonamputated) 1-point function. The rest of the section is devoted to the
procedure for attaching the retarded propagators, needed in the Schwinger
formalism \cite{Schw}, to convert amputated 1-point functions into their
nonamputated cognates. Although an exact solution is obtained we specialize 
it, at the very end of the section, to leading order in the slow roll 
approximation.

Both the initial state and the evolution equations are homogeneous and 
isotropic. It follows that the expectation values of the (general) 
pseudo-graviton field and the scalar can be expressed in terms of three
functions of $\eta$. It will simplify some of the later formulae if we choose
to think of these quantities as functions of the co-moving time $t$ of the
perturbative background, even though we still are expressing them in conformal
coordinates:\footnote{Recall that the relation between $t$ and 
$\eta$ is:
\begin{eqnarray*}
dt = \Omega(\eta) d\eta \Longleftrightarrow {dt \over a_0(t)} = d\eta \; 
\end{eqnarray*}}
\begin{eqnarray}
\left\langle 0 \left\vert \kappa \psi_{\mu\nu}(\eta,{\vec x}) \right\vert 
0 \right\rangle & = & A(t) {\overline \eta}_{\mu\nu} + C(t) t_{\mu} t_{\nu} 
\; , \label{eq:unamppsi} \\
\left\langle 0 \left\vert \kappa \phi(\eta,{\vec x}) \right\vert 0 
\right\rangle & = & D(t) \; . \label{eq:unampphi} 
\end{eqnarray}
Note that we cannot assume $C = - A$ since the expectation value may not be
conformal in the perturbative coordinate system. None of these quantities is
itself physical but they can be combined to produce observables. We first 
construct the invariant element to infer the true scale factor and co-moving
time ${\overline t}$ of the expectation value of the metric:
\begin{eqnarray}
-d{\overline t}^2 + a^2({\overline t}) d{\vec x} \cdot d{\vec x} & = & \Omega^2 
\left\{- \left(1 - C \right) d\eta^2 + \left(1 + A\right) d{\vec x} \cdot 
d{\vec x}\right\} \; ,\\
& = & -[1-C(t)] dt^2 + a^2_0(t) [1 + A(t)] d{\vec x} \cdot d{\vec x} \; .
\end{eqnarray}
One physical observable is the effective Hubble constant expressed as a
function of the co-moving time ${\overline t}$:
\begin{equation}
H_{\rm eff}({\overline t}) \equiv {d \over d{\overline t}} \ln[a({ \overline 
t})] = {1 \over \sqrt{1 - C(t)}} \left\{H(t) + \frac12 {\dot{A}(t) \over 1 + 
A(t)}\right\} \; . \label{eq:Heff}
\end{equation}
If the scalar can be measured then its expectation value is also an
observable when expressed as a function of the co-moving time ${\overline t}$.
We shall call this variable $\Phi({\overline t})$:
\begin{equation}
\Phi({\overline t}) \equiv \varphi_0(t) + {1 \over \kappa} D(t) \; . 
\label{eq:Phi}
\end{equation}

What we actually compute are not the expectation values of $\kappa \psi_{\mu
\nu}$ and $\kappa \phi$ but rather the {\it amputated} expectation values with 
the external propagators removed. We will use Greek letters to denote the three
functions of $t$ which describe these amputated quantities:
\begin{eqnarray}
\alpha(t) {\overline \eta}_{\mu\nu} +\gamma(t) t_{\mu} t_{\nu} & \equiv &
D_{\mu\nu}^{~~\rho\sigma} \langle 0 \vert \kappa \psi_{\rho\sigma}(\eta,{\vec
x}) \vert 0 \rangle \nonumber \\
& & \hskip 1cm \mbox{} - \kappa \Omega^2 \varphi_0^{\prime\prime} t_{\mu} 
t_{\nu} \langle 0 \vert \kappa \phi(\eta,{\vec x}) \vert 0 \rangle \; , 
\label{eq:ampmetric} \\
\delta(t) & \equiv & -\kappa \Omega^2 \varphi_0^{\prime\prime} t^{\rho} 
t^{\sigma} \langle 0 \vert \kappa \psi_{\rho\sigma}(\eta,{\vec x}) \vert 0
\rangle \nonumber \\
& & \mbox{} + \Omega \left(\partial^2 + \frac14 \kappa^2 
{\varphi_0^{\prime}}^2 + {\varphi_0^{\prime\prime\prime} \over \varphi_0^{
\prime}}\right) \Omega \langle 0 \vert \kappa \phi(\eta,{\vec x})\vert 0 
\rangle \; . \qquad \label{eq:ampscalar}
\end{eqnarray}
Contracting with the kinetic operator (\ref{eq:kinetic}) and isolating 
distinct tensor components gives three relations:
\begin{eqnarray}
\alpha & = & -\frac14 {\rm D}_A (A - C) \; , \label{eq:arel} \\
\gamma & = & \frac34 {\rm D}_A (A - C) + {\rm D}_B C - \kappa \Omega^2 
\varphi_0^{\prime\prime} D \; , \label{eq:crel} \\
\delta & = & -\kappa \Omega^2 \varphi_0^{\prime\prime} C + \Omega \left[-{d^2 
\over d\eta^2} + \frac14 \kappa^2 {\varphi_0^{\prime\prime}}^2 + {\varphi_0^{
\prime\prime\prime} \over \varphi_0^{\prime}}\right] \Omega D \; . 
\label{eq:drel}
\end{eqnarray}
Since it is from $A$ and $C$ that $H_{\rm eff}$ is constructed, we must invert
these relations.

We employ the Schwinger formalism \cite{Schw} in order to get true expectation 
values rather than in-out matrix elements. An important feature of this 
formalism is that external legs are retarded propagators. This means that the 
coupled differential equations in (\ref{eq:arel}-\ref{eq:drel}) must be 
inverted using retarded boundary conditions:
\begin{eqnarray}
0 & = & A(0) = C(0) = D(0) \; ,\\
0 & = & \dot{A}(0) = \dot{C}(0) = \dot{D}(0) \; ,
\end{eqnarray}
Now it happens that every differential equation we have to solve can be cast in
the form:
\begin{equation}
{\cal D} f(t) \equiv \left(-{d^2 \over d\eta^2} + {\theta^{\prime\prime} 
\over \theta}\right) f(t) = g(t) \; . \label{eq:generalD}
\end{equation}
This is fortunate because the retarded solution can be simply expressed as a
double integral:
\begin{equation}
f(t) = {\cal D}^{-1}(g) \equiv -\theta(\eta) \int_0^t dt_1 a_0^{-1}(t_1)
\theta^{-2}(\eta_1) \int_0^{t_1} dt_2 a_0^{-1}(t_2) \theta(\eta_2) g(t_2) 
\;. \label{eq:geninv}
\end{equation}

It simplifies the algebra somewhat to multiply the amputated quantities by
$\Omega^{-1}$ and their unamputated descendants by $\Omega$. We denote the
rescaled variables by a tilde:
\begin{eqnarray}
& & {\widetilde A} \equiv \Omega A \qquad , \qquad {\widetilde C} \equiv 
\Omega C \qquad , \qquad {\widetilde D} \equiv \Omega D \; ,\\
& & {\widetilde \alpha} \equiv \Omega^{-1} \alpha \qquad ,\qquad {\widetilde 
\gamma} \equiv \Omega^{-1} \gamma \qquad , \qquad {\widetilde \delta} \equiv 
\Omega^{-1} \delta \; . \label{eq:amptilde}
\end{eqnarray}
In this notation the equations we must invert are:
\begin{eqnarray}
{\widetilde \alpha} & = & -\frac14 {\cal D}_A ({\widetilde A} - {\widetilde C}) 
\; , \label{eq:atilde} \\
{\widetilde \gamma} & = & \frac34 {\cal D}_A ({\widetilde A}- {\widetilde C}) + 
{\cal D}_B {\widetilde C} - \kappa \varphi_0^{\prime\prime} {\widetilde D} 
\; , \label{eq:ctilde} \\
{\widetilde \delta} & = & -\kappa \varphi_0^{\prime\prime} {\widetilde C} + 
\left[-{d^2 \over d\eta^2} + \frac14 \kappa^2 {\varphi_0^{\prime\prime}}^2 + 
{\varphi_0^{\prime\prime\prime} \over \varphi_0^{\prime}}\right] {\widetilde D}
\; , \label{eq:dtilde} 
\end{eqnarray}
where ${\cal D}_A$ and ${\cal D}_B$ have the form (\ref{eq:generalD}) with 
$\theta_A = \Omega$ and $\theta_B = \Omega^{-1}$. Equation (\ref{eq:atilde}) 
implies:
\begin{equation}
{\widetilde A} = {\widetilde C} + {\cal D}_A^{-1}\left(-4 {\widetilde \alpha}
\right) \; .
\end{equation}
Substituting this into (\ref{eq:ctilde}) gives:
\begin{equation}
3 {\widetilde \alpha} + {\widetilde \gamma}= {\cal D}_B {\widetilde C} - \kappa 
\varphi_0^{\prime\prime} {\widetilde D} \; , \label{eq:newctilde}
\end{equation}
which, with (\ref{eq:dtilde}), is similar to the coupled $z$--$f$ system of 
Section 4. Paralleling the analysis of that section we differentiate 
(\ref{eq:newctilde}) and add it to $\frac12 \kappa \varphi_0^{\prime}$ times 
(\ref{eq:dtilde}) to obtain:
\begin{equation}
{\cal D}_B \left({\widetilde C}^{\prime} + \frac12 \kappa \varphi_0^{\prime}
{\widetilde D}\right) = 3 {\widetilde \alpha}^{\prime} + {\widetilde \gamma}^{
\prime} + \frac12 \kappa \varphi_0^{\prime} {\widetilde \delta} \; .
\end{equation}
Differentiating (\ref{eq:dtilde}) and adding it to $\frac12 \kappa \varphi_0^{
\prime}$ times (\ref{eq:newctilde}) minus $\varphi_0^{\prime\prime}/\varphi_0^{
\prime}$ times (\ref{eq:dtilde}) gives:
\begin{equation}
{\cal D}_C \left(\frac12 \kappa \varphi_0^{\prime} {\widetilde C} + {\widetilde
D}^{\prime} - {\varphi_0^{\prime\prime} \over \varphi_0^{\prime}} {\widetilde
D}\right) = \frac12 \kappa \varphi_0^{\prime}\left(3 {\widetilde \alpha} + 
{\widetilde \gamma}\right) + {\widetilde \delta}^{\prime} - {\varphi_0^{\prime
\prime} \over \varphi_0^{\prime}} {\widetilde \delta} \; ,
\end{equation}
where ${\cal D}_C$ has the form (\ref{eq:generalD}) with $\theta_C = -2 
\kappa^{-1} \Omega^{-2} \Omega^{\prime}/\varphi_0^{\prime}$.

Of course we can invert the differential operators in the last two equations:
\begin{eqnarray}
{\widetilde C}^{\prime} + \frac12 \kappa \varphi_0^{\prime} {\widetilde D} &= & 
{\cal D}_B^{-1}\left(3 {\widetilde \alpha}^{\prime} + {\widetilde \gamma}^{
\prime} + \frac12 \kappa \varphi_0^{\prime} {\widetilde \delta}\right) \; , 
\label{eq:D_B} \\
\frac12 \kappa \varphi_0^{\prime} {\widetilde C} + {\widetilde D}^{\prime} -
{\varphi_0^{\prime\prime} \over \varphi_0^{\prime}} {\widetilde D} & = &
{\cal D}_C^{-1} \left(\frac12 \kappa \varphi_0^{\prime} (3 {\widetilde \alpha}+
{\widetilde \gamma}) + {\widetilde \delta}^{\prime} - {\varphi_0^{\prime\prime}
\over \varphi_0^{\prime}} {\widetilde \delta}\right) \; , \label{eq:D_C}
\end{eqnarray}
but this still leaves derivatives on ${\widetilde C}$ and ${\widetilde D}$. 
These derivatives cannot be removed as we did in Section 4 because the 
Laplacian vanishes for spatially homogeneous functions. What we must do 
instead is to divide (\ref{eq:newctilde}) by $\varphi_0^{\prime}$ and add 
it to $2\varphi_0^{\prime\prime}/{\varphi_0^{\prime}}^2$ times (\ref{eq:D_B}):
\begin{equation}
{\cal D}_C \left({{\widetilde C} \over \varphi_0^{\prime}}\right) = {3 
{\widetilde \alpha} + {\widetilde \gamma} \over \varphi_0^{\prime}} + 2 
{\varphi_0^{\prime\prime} \over {\varphi_0^{\prime}}^2} {\cal D}_B^{-1} \left(3
{\widetilde \alpha}^{\prime} + {\widetilde \gamma}^{\prime} + \frac12 \kappa 
\varphi_0^{\prime} {\widetilde \delta}\right) \; .
\end{equation}
Dividing (\ref{eq:dtilde}) by $\varphi_0^{\prime}$ and adding it to $2
\varphi_0^{\prime\prime}/{\varphi_0^{\prime}}^2$ times (\ref{eq:D_C}) gives a
similar relation for ${\widetilde D}$:
\begin{equation}
{\cal D}_B \left({{\widetilde D} \over \varphi_0^{\prime}}\right) = 
{{\widetilde \delta} \over \varphi_0^{\prime}} + 2 {\varphi_0^{\prime\prime} 
\over {\varphi_0^{\prime}}^2} {\cal D}_C^{-1} \left(\frac12 \kappa \varphi_0^{
\prime} (3 {\widetilde \alpha} + {\widetilde \gamma}) + {\widetilde \delta}^{
\prime} - {\varphi_0^{\prime\prime} \over \varphi_0^{\prime}} {\widetilde 
\delta}\right) \; .
\end{equation}

Putting everything together gives the following solution for the unamputated
coefficient functions:
\begin{eqnarray}
A & = & \Omega^{-1} {\cal D}_A^{-1}(-4 {\widetilde \alpha}) \nonumber \\
& & \hskip 1cm \mbox{}+ {\varphi_0^{\prime} 
\over \Omega} {\cal D}_C^{-1} \left\{ {3 {\widetilde \alpha} + {\widetilde 
\gamma} \over \varphi_0^{\prime}} + 2 {\varphi_0^{\prime\prime} \over 
{\varphi_0^{\prime}}^2} {\cal D}_B^{-1} \left(3 {\widetilde \alpha}^{\prime} + 
{\widetilde \gamma}^{\prime} + \frac12 \kappa \varphi_0^{\prime} {\widetilde 
\delta}\right)\right\} \; , \label{eq:retA} \qquad \\
C & = & {\varphi_0^{\prime} \over \Omega} {\cal D}_C^{-1} \left\{ {3 
{\widetilde \alpha} + {\widetilde \gamma} \over \varphi_0^{\prime}} + 2 
{\varphi_0^{\prime\prime} \over {\varphi_0^{\prime}}^2} {\cal D}_B^{-1} \left(3
{\widetilde \alpha}^{\prime} + {\widetilde \gamma}^{\prime} + \frac12 \kappa 
\varphi_0^{\prime} {\widetilde \delta}\right)\right\} \; , \label{eq:retC} \\
D & = & {\varphi_0^{\prime} \over \Omega} {\cal D}_B^{-1} \left\{{{\widetilde 
\delta} \over \varphi_0^{\prime}} + 2 {\varphi_0^{\prime\prime} \over 
{\varphi_0^{\prime}}^2} {\cal D}_C^{-1} \left(\frac12 \kappa \varphi_0^{\prime}
(3 {\widetilde \alpha} + {\widetilde \gamma}) + {\widetilde \delta}^{\prime} - 
{\varphi_0^{\prime\prime} \over \varphi_0^{\prime}} {\widetilde \delta}\right) 
\right\} \; . \label{eq:retD}
\end{eqnarray}
Recall that ${\widetilde \alpha} \equiv \Omega^{-1} \alpha$, ${\widetilde 
\gamma} \equiv \Omega^{-1} \gamma$ and ${\widetilde \delta} \equiv \Omega^{-1}
\delta$.  The various inverse differential operators are defined by 
(\ref{eq:geninv}) with the following assignments for $\theta(\eta)$:
\begin{equation}
\theta_A = \Omega \qquad , \qquad \theta_B = \Omega^{-1} \qquad , \qquad
\theta_C = -{2 \over \kappa} {\Omega^{\prime} \over \Omega^2 
\varphi_0^{\prime}} \; .
\end{equation}

For the potential $V(\varphi) = \frac12 m^2 \varphi^2$ it happens that the 
leading order results for the amputated 1-point functions consist of sums of
terms with the general form:
\begin{eqnarray}
\alpha(t) & = & \alpha_N H^N(t) a_0^4(t) + \dots \; ,\\
\gamma(t) & = & \gamma_N H^N(t) a_0^4(t) + \dots \; ,\\
\delta(t) & = & \delta_N H^N(t) \left({\sqrt{-\dot{H}} \over H}\right) a_0^4(t)
+ \dots \; .
\end{eqnarray}
The coefficients $\alpha_N$, $\gamma_N$ and $\delta_N$ are constants. When the 
slow roll expansions of Section 2 are applied to the various integrations and 
differentiations in our formulae for $A$, $C$ and $D$, the following leading 
order results emerge:
\begin{eqnarray}
A_N(t) & = & {4 \alpha_N \over 3 N} \left({H_I^N - H^N(t) \over H^2(t)}\right) 
\left({H^2 \over -\dot{H}}\right) + \dots \; ,\\
C_N(t) & = & \left[\left({3 \alpha_N + \gamma_N \over 2}\right) - \left({3 
\alpha_N + \gamma_N + \delta_N \over 3 N}\right)\right] \left({H_I^N- H^N(t) 
\over H^2(t)}\right)+\dots \; ,\\
D_N(t) & = & - \left({3 \alpha_N + \gamma_N + \delta_N \over 3 N}\right) 
\left({H_I^N- H^N(t) \over H^2(t)}\right) \left({H \over\sqrt{-\dot{H}}}\right)
+ \dots \; .
\end{eqnarray}
Here $H_I \equiv H(0)$ is the Hubble constant at the beginning of inflation.
These results imply the following leading order shift in co-moving time:
\begin{eqnarray}
({\overline t} - t)_N &= &-\frac12 \left[\frac32 \alpha_N - \left({3 \alpha_N + 
\gamma_N + \delta_N \over 3 N}\right)\right] \left({H_I^N \over H^2} \right. 
\nonumber \\
& & \qquad \qquad \left. - {N \over N-1} {H_I^{N-1} \over H} + {H^{N-2} \over 
N-1} \right) \left({H \over \sqrt{-\dot{H}}}\right) + \dots \; . \qquad
\end{eqnarray}
To leading order the proportional shift in the two observables is:
\begin{eqnarray}
& & \left({H_{\rm eff}- H \over H}\right)_N = \frac23 \alpha_N H^{N-2} + \left[
\left({3\alpha_N+ \gamma_N \over 4}\right) \left({N \over N-1}\right)\right. 
\nonumber \\
& & \qquad \qquad \left. - \left({3 \alpha_N+\gamma_N + \delta_N \over 6 (N-1)}
\right) \right] \left({H_I^{N-1} - H^{N-1} \over H}\right) +
\dots \; , \qquad \qquad \label{eq:HeffN}
\end{eqnarray}
\begin{eqnarray}
\left({\Phi - \varphi_0 \over \varphi_0}\right)_N & = & -\left({3 \alpha_N +
\gamma_N \over 4}\right) \left[{H_I^N \over H^2} - {N \over N-1} {H_I^{N-1} 
\over H} + {H^{N-2} \over N-1}\right] \nonumber \\
& & \mbox{} - \left({3 \alpha_N + \gamma_N + \delta_N \over 6 (N-1)}\right) 
\left[H^{-1} H_I^{N-1} - H^{N-2}\right] + \dots \; . \qquad \label{eq:PhiN}
\end{eqnarray}
To obtain the full shift one sums the contributions for various different 
values of $N$.

\section{Infrared parts of propagators}

This section deals with a very important omission in Section 4. Although we 
were able there to express the various propagators as mode sums for an 
arbitrary scalar potential, we do not possess the corresponding mode {\it 
functions} $Q_I(\eta,k)$ for a general potential. This is a standard problem in
the theory of cosmological perturbations \cite{MFB,LL} and we solve it in the
standard way: by developing series solutions for the ultraviolet (early time)
and infrared (late time) regimes. The normalization for the ultraviolet 
expansion derives from the flat space limit. We normalize the infrared
expansion by matching its leading term with that of the ultraviolet expansion
at the time when the physical wavelength of each mode is just redshifting 
beyond the Hubble radius. (This is the chief approximation of the paper.) One 
then defines the ``infrared part'' of each propagator as that obtained from the
leading order term of the infrared expansion. We report explicit results to
leading order in the slow roll approximation.

Recall from Section 4 that we have three kinds of plane wave mode solutions 
$Q_I(\eta,k)$. They obey the equation:
\begin{equation}
{\cal D}_I Q_I(\eta,k) \equiv - \left\{{d \over d\eta^2} + k^2 - {\theta_I^{
\prime\prime}(\eta) \over \theta_I(\eta)}\right\} Q_I(\eta,k) = 0 \; ,
\label{eq:mode}
\end{equation}
where the $\theta_I(\eta)$'s are:
\begin{equation}
\theta_A \equiv \Omega \qquad , \qquad \theta_B \equiv \Omega^{-1} \qquad , 
\qquad \theta_C \equiv -{2 \over \kappa} {\Omega^{\prime} \over \Omega^2
\varphi_0^{\prime}} = a_0^{-1} {H \over \sqrt{-\dot{H}}} \; .
\end{equation}
Of course (\ref{eq:mode}) does not completely define the modes because
there are two linearly independent solutions. We define $Q_I(\eta,k)$ as
the solution of (\ref{eq:mode}) which is canonically normalized:
\begin{equation}
Q(\eta,k) Q^{*\prime}(\eta,k) - Q^{\prime}(\eta,k) Q^*(\eta,k) = i \; ,
\label{eq:Wronskian}
\end{equation}
and descends by perturbative iteration from the negative frequency solution 
of the far ultraviolet.

The far ultraviolet is defined by $k^2 \gg \theta^{\prime\prime}/\theta$. At 
fixed $k$ this condition will also be realized, in all models of inflation,
as the conformal time approaches negative infinity. In the ultraviolet regime
we build up normalized solutions by iterating the following equation:
\begin{equation}
Q_I(\eta,k) = {1 \over \sqrt{2 k}} e^{-i k \eta} + \int_{-\infty}^{\eta}
d{\overline \eta} {1 \over k} \sin\left[k (\eta - {\overline \eta})\right]
{\theta_I^{\prime\prime}({\overline \eta}) \over \theta_I({\overline \eta})}
Q_I({\overline \eta},k) \; .
\end{equation}
The result is a series in inverse powers of $k$. These solutions are obviously
negative frequency in the far ultraviolet. Their Wronskian (\ref{eq:Wronskian})
is constant as a simple consequence of the mode equation (\ref{eq:mode}) while
its actual value derives from the fact that $\theta^{\prime\prime}/\theta$ 
vanishes as the conformal time approaches negative infinity.

The far infrared is defined by $k^2 \ll \theta^{\prime\prime}/\theta$. At 
fixed $k$ this condition will also be realized, in all models of inflation, as
the conformal time approaches zero from below. One can find explicit solutions 
in the limit that the $k^2$ term is neglected. The first one has the same form
for $I= A,B,C$:
\begin{equation}
Q_{10,I}(\eta) \equiv \theta_I(\eta) \; .
\end{equation}
The second is an integral whose convergence (for models of inflation) requires 
different limits for $I=A$:
\begin{equation}
Q_{20,A}(\eta) \equiv - \theta_I(\eta) \int_{\eta}^0 {d{\overline 
\eta} \over \theta_A^2({\overline \eta})} \; , 
\end{equation}
and $I = B,C$:
\begin{equation}
Q_{20,I}(\eta) \equiv \theta_I(\eta) \int_{\infty}^{\eta} {d{\overline 
\eta} \over \theta_I^2({\overline \eta})} \qquad (I = B,C) \; .
\end{equation}
When $k^2$ is small but not zero one can build up solutions which descend 
from $Q_{i0,I}(\eta)$ by iterating with the appropriate Green's function:
\begin{equation}
Q_{i,I}(\eta,k) = Q_{i0,I}(\eta) - k^2 \int_{-\infty}^0 d{\overline \eta}
\; G_{\rm app}(\eta,{\overline \eta}) Q_{i,I}(\eta,k) \qquad (i=1,2) \; , 
\end{equation}
This obviously gives a series of increasing powers of $k^2$. Here the 
``appropriate'' Green's function is chosen to make the integral converge. The 
four possibilities are:
\begin{eqnarray}
G_{\rm adv}(\eta,{\overline \eta}) & = & + \theta({\overline \eta} - \eta) 
Q_{10}(\eta) Q_{20}({\overline \eta}) - \theta({\overline \eta} - \eta) 
Q_{20}(\eta) Q_{10}({\overline \eta}) \; ,\\
G_{12}(\eta,{\overline \eta}) & = & - \theta(\eta - {\overline \eta}) 
Q_{10}(\eta) Q_{20}({\overline \eta}) - \theta({\overline \eta} - \eta)
Q_{20}(\eta) Q_{10}({\overline \eta}) \; ,\\
G_{21}(\eta,{\overline \eta}) & = & + \theta(\eta - {\overline \eta}) 
Q_{20}(\eta) Q_{10}({\overline \eta}) + \theta({\overline \eta} - \eta)
Q_{10}(\eta) Q_{20}({\overline \eta}) \; ,\\
G_{\rm ret}(\eta,{\overline \eta}) & = & + \theta(\eta - {\overline \eta}) 
Q_{20}(\eta) Q_{10}({\overline \eta}) - \theta(\eta - {\overline \eta}) 
Q_{10}(\eta) Q_{20}({\overline \eta}) \; ,
\end{eqnarray}
and it should be noted that one may have to switch from one to another midway
through the iteration process.

Since the full infrared solutions, $Q_{1,I}(\eta,k)$ and $Q_{2,I}(\eta,k)$, 
span the space of solutions to ${\cal D}_I = 0$, it must be possible to 
express the ultraviolet solutions as linear combinations:
\begin{equation}
Q_I(\eta,k) = q_1 Q_{1,I}(\eta,k) + q_2 Q_{2,I}(\eta,k) \; .
\end{equation}
If we had the full solutions it would be straightforward to determine the
combination coefficients:
\begin{eqnarray}
q_1 & = & {Q_{2,I}^{\prime}(\eta,k) Q_I(\eta,k) - Q_{2,I}(\eta,k) 
Q_I^{\prime}(\eta,k) \over Q_{1,I}(\eta,k) Q_{2,I}^{\prime}(\eta,k) - 
Q_{2,I}(\eta,k) Q_{1,I}^{\prime}(\eta,k)} \; , \\
q_2 & = & {-Q_{1,I}^{\prime}(\eta,k) Q_I(\eta,k) + Q_{1,I}(\eta,k) 
Q_I^{\prime}(\eta,k) \over Q_{1,I}(\eta,k) Q_{2,I}^{\prime}(\eta,k) - 
Q_{2,I}(\eta,k) Q_{1,I}^{\prime}(\eta,k)} \; ,
\end{eqnarray}
where any conformal time $\eta$ could be chosen. 

For most backgrounds we do not possess the full solutions --- either in the 
ultraviolet or the infrared. However, it happens that one of the zeroth order 
infrared solutions --- either $Q_{10,I}(\eta)$ or $Q_{20,I}(\eta)$ --- 
dominates the other and all corrections as the conformal time approaches zero 
from below. The standard approximation \cite{MFB,LL} is to match this solution 
with the zeroth order ultraviolet solution at the horizon crossing time 
$\eta_*$, whose defining condition is:
\begin{equation}
k = H_* \Omega(\eta_*) \; . \label{eq:horizon}
\end{equation}
Then the behavior of the modes in the far infrared can be approximated as
follows:
\begin{equation}
Q_I(\eta,k) \longrightarrow {Q_{i0,I}(\eta) \over Q_{i0,I}(\eta_*)} {e^{-i k 
\eta_*} \over \sqrt{2 k}} \; ,
\end{equation}
where $i$ is either 1 or 2, depending upon which of the zeroth order infrared
solutions dominates for the $I$ mode.

For $I=A$ it is $Q_{10,A}(\eta) = \Omega(\eta)$ that dominates at late times.
We can therefore write:
\begin{equation}
Q_A(\eta,k) \longrightarrow \Omega(\eta) \cdot {H_* \over k} {e^{-i k \eta_*}
\over \sqrt{2 k}} \; . \label{eq:Amode}
\end{equation}
For $I=B$ $Q_{20,B} = \Omega^{-1}$ becomes irrelevant at late times. The
dominant solution is:
\begin{equation}
Q_{20,B}(\eta) = \Omega^{-1} \int_{-\infty}^{\eta} d{\overline \eta} \Omega^2(
{\overline \eta}) = {1 \over a_0(t)} \int_{-\infty}^t dt' a_0(t') \approx {1 
\over H(t)} \; .
\end{equation}
We can therefore approximate the $B$ modes as follows:
\begin{equation}
Q_B(\eta,k) \longrightarrow {1 \over H(t)} \cdot H_* {e^{-i k \eta_*} \over
\sqrt{2 k}} \; . \label{eq:Bmode}
\end{equation}
The case of $I=C$ requires a more extensive analysis. The first solution is 
down by an inverse scale factor but enhanced by the inverse of a slow roll
parameter:
\begin{equation}
Q_{10,C} = a_0^{-1} {H \over \sqrt{-\dot{H}}} \; .
\end{equation}
The second solution has to be re-expressed several times before it can be
recognized as the dominant one:
\begin{eqnarray}
Q_{20,C} & = & a_0^{-1} {H \over \sqrt{-\dot{H}}} \int_{-\infty}^t dt' a_0(t')
{d \over dt'} \left({1 \over H(t')}\right) \; ,\\
& = & {1 \over \sqrt{-\dot{H}}} {d \over dt} \left\{ {1 \over a_0(t)} \int_{-
\infty}^t dt' a_0(t') \right\} \; , \\
& \approx & {\sqrt{-\dot{H}} \over H^2} \; .
\end{eqnarray}
After any significant amount of inflation the inverse scale factor is much 
smaller than $-\dot{H}/H^2$, so we can approximate the $C$ modes as:
\begin{equation}
Q_C(\eta,k) \longrightarrow {\sqrt{-\dot{H}(t)} \over H^2(t)} \cdot {H^2_*
\over \sqrt{-\dot{H}_*}} {e^{-i k \eta_*} \over \sqrt{2 k}} \; . 
\label{eq:Cmode}
\end{equation}

Although the approximations (\ref{eq:Amode},\ref{eq:Bmode},\ref{eq:Cmode}) we
have just made may seem grotesque they are intimately related to the physics 
of superadiabatic amplification \cite{Gris}. It is this phenomenon's vast 
enhancement of the usual 0-point energy which causes one of the zeroth order 
infrared solutions to dominate at late times. These approximations therefore 
isolate precisely the leading late time infrared effect we wish to study. In 
fact this is all that {\it can} be reliably studied using quantum general 
relativity. The ultraviolet regime, which these approximations fail to 
capture, cannot in any case be described perturbatively by quantum general 
relativity.

What remains is to implement the infrared approximations (\ref{eq:Amode},
\ref{eq:Bmode},\ref{eq:Cmode}) in the various propagator mode sums derived in
Section 4. Since we are only computing the amputated 1-point function to one 
loop order, these propagators are all coincident. They may, however, bear 
derivatives. Since space derivatives add factors of $k$, which are small in 
the infrared, we need only consider time derivatives. We shall therefore set
${\Delta x} = 0$ but keep the two times nonzero. With these conventions all
the propagators can be described in the standard form:
\begin{equation}
i \Delta(x;x') \longrightarrow f(t) \cdot g(t') \cdot \int dk {H_*^2 \over 2 k} 
\cdot h(k) \; ,
\end{equation}
where it should be noted that $H_*$ and $\dot{H}_*$ are functions of the
co-moving wave number $k$, determined by the horizon crossing condition
(\ref{eq:horizon}).

From (\ref{eq:Amode}) and the mode sums (\ref{eq:bigprop},\ref{eq:smallprop})
which define it we see that the infrared part of $i \Delta_A(x:x')$ approaches
a constant:
\begin{equation}
i \Delta_A(x;x') \longrightarrow 1 \cdot 1 \cdot {1 \over 2 \pi^2} 
\int dk {H_*^2 \over 2 k} \cdot 1 \; . \label{eq:SRA}
\end{equation}
The behavior of the $B$ mode (\ref{eq:Bmode}) and the mode sums 
(\ref{eq:bigprop},\ref{eq:smallprop}) which define $i \Delta_B(x;x')$ show that
its infrared part actually falls off:
\begin{equation}
i \Delta_B(x;x') \longrightarrow {1 \over a_0(t) H(t)} \cdot {1 \over a_0(t')
H(t')} \cdot {1 \over 2 \pi^2} \int dk {H_*^2 \over 2 k} \cdot k^2 \; .
\label{eq:SRB}
\end{equation}
Since the momentum integral is dominated by the ultraviolet, rather than the
infrared, we conclude that the infrared part of this propagator is zero. The
$\psi_{00}$ propagator involves $B$ modes (\ref{eq:Bmode}) and $C$ modes
(\ref{eq:Cmode}). From its defining relations (\ref{eq:alphaprop},
\ref{eq:smallprop}) we determine its infrared part to be:
\begin{equation}
i \Delta_{\alpha}(x;x') \longrightarrow {-\dot{H}(t) \over H^2(t)} \cdot
{-\dot{H}(t') \over H^2(t')} \cdot {1 \over 2 \pi^2} \int dk {H_*^2 \over
2 k} \cdot {H_*^2 \over -\dot{H}_*} \; . \label{eq:SRalpha}
\end{equation}
The mixed propagator is defined by relations 
(\ref{eq:betaprop},\ref{eq:smallprop}). Again applying the infrared mode 
approximations (\ref{eq:Bmode},\ref{eq:Cmode}) gives:
\begin{equation}
i \Delta_{\beta}(x;x') \longrightarrow {-\dot{H}(t) \over H^2(t)} \cdot
{\sqrt{-\dot{H}(t')} \over H(t')} \cdot {1 \over 2 \pi^2} \int dk {H_*^2 \over
2 k} \cdot {H_*^2 \over -\dot{H}_*} \; . \label{eq:SRbeta}
\end{equation}
The same infrared limits, applied to its defining relations 
(\ref{eq:gammaprop},\ref{eq:smallprop}), reduce the scalar propagator to:
\begin{equation}
i \Delta_{\gamma}(x;x') \longrightarrow {\sqrt{-\dot{H}(t)} \over H(t)} \cdot
{\sqrt{-\dot{H}(t')} \over H(t')} \cdot {1 \over 2 \pi^2} \int dk {H_*^2 \over
2 k} \cdot {H_*^2 \over -\dot{H}_*} \; . \label{eq:SRgamma}
\end{equation}
By itself it is the strongest but one must allow for the effect of factors
and derivatives from the interaction vertex.

Since modes only become infrared after horizon crossing, the momentum 
integrations are cut off at $k = H(t) a_0(t)$. The integral can be evaluated by
first changing variables from $k$ to the horizon crossing time $t_*$:
\begin{equation}
dk \approx H^2(t_*) a_0(t_*) dt_* \; ,
\end{equation}
and then employing the slow roll expansions of Section 2:
\begin{eqnarray}
{1 \over 2 \pi^2} \int dk {H_*^2 \over 2 k} \left({H_*^2 \over - \dot{H}_*}
\right) & \approx & {1 \over 4 \pi^2} \int_0^t dt_* {H^5(t_*) \over -\dot{H}}
\; , \\
& \approx & {1 \over 24 \pi^2} \left({H_I^6 - H^6(t) \over \dot{H}^2}\right)
\; .
\end{eqnarray}
Note that $\dot{H}$ is approximately constant for the quadratic potential 
considered by Mukhanov, Abramo and Brandenberger \cite{Mukh}.

\section{Amputated 1-point functions}

The purpose of this section is first to obtain one loop results for the three 
amputated 1-point functions (\ref{eq:ampmetric}-\ref{eq:ampscalar}) defined in 
Section 5. We then exploit the technology of Section 5 to compute the two 
observables $H{\rm eff}({\overline t})$ (\ref{eq:Heff})and $\Phi({\overline 
t})$ (\ref{eq:Phi}). We begin by explaining how cubic interactions are used to 
compute the amputated 1-point functions.

At one loop order the amputated 1-point functions consist basically of 
coincident propagators contracted into cubic interaction vertices. In addition 
to the usual factor $i$ there is an $i$ from the kinetic operator acting on the
external propagator. There is also an extra factor of $\kappa$ from the fact
that we define the 1-point functions (\ref{eq:unamppsi}-\ref{eq:unampphi}) as
$\langle 0 \vert \kappa \psi_{\mu\nu} \vert 0 \rangle$ and $\langle 0 \vert
\kappa \phi \vert 0 \rangle$.

As an example let us consider the interaction $-\frac14 \kappa m^2 \Omega^4 
\phi^2 \psi$, which is one of the many cubic terms descending from the scalar 
mass. The external line can attach to any of the three quantum fields: $\phi^2 
\psi_{\mu\nu} \eta^{\mu\nu}$. When it attaches to one of the two scalar fields 
$\delta(t)$ receives the following contribution:
\begin{eqnarray}
\lefteqn{\left\{i \cdot \kappa \cdot -\frac{i}4 \kappa m^2 \Omega^4 \cdot 2 
\cdot \langle 0 \vert T\left[ \phi(x) \psi(x')\right] \vert 0 \rangle\right\}_{
x' = x}} \nonumber \\
& & \qquad \qquad=\left\{ \frac12 \kappa^2 m^2 \Omega^4 i{\Delta_{\beta}(x;x')}
({\overline \eta}_{\mu\nu} + t_{\mu} t_{\nu})\eta^{\mu\nu}\right\}_{x'=x} 
\; ,\\
& & \qquad \qquad = \left\{\kappa^2 m^2 \Omega^4 i{\Delta_{\beta}(x;x')}
\right\}_{x'=x} \; .
\end{eqnarray}
Most of the coincindence limit is ultraviolet nonsense which quantum general
relativity cannot be trusted to treat correctly and which must in any case have
been subtracted off in order for inflation to begin in the first place. The 
time dependent, physically significant part comes from the superadiabatically
amplified infrared modes. From equation (\ref{eq:SRbeta}) we see that the 
leading effect from these is:
\begin{eqnarray}
\kappa^2 m^2 \Omega^4 i \Delta_{\beta}(x;x) & \stackrel{IR}{\longrightarrow} & 
-3 \kappa^2 \dot{H} a^4_0(t) \cdot {(-\dot{H})^{\frac32} \over H^3(t)} \cdot 
{1 \over 24 \pi^2} \left({H_I^6 - H^6(t) \over \dot{H}^2}\right) 
\; , \nonumber \\
& = & {\kappa^2 \over 8 \pi^2} \left({H_I^6 - H^6(t) \over H^2(t)}\right)
\left({\sqrt{-\dot{H}} \over H(t)}\right) a_0^4(t) \; .
\end{eqnarray}
In the notation used at the end of Section 5 there are contributions for $N=-2$
and $N=+4$ with coefficients of $\kappa^2/(8\pi^2)$ times $+H^6_I$ and $-1$,
respectively.

\begin{table}

\vbox{\tabskip=0pt \offinterlineskip
\def\tablerule{\noalign{\hrule}}
\halign to460pt {\strut#& \vrule#\tabskip=1em plus2em& 
\hfil#& \vrule#& \hfil#\hfil& \vrule#& \hfil#& \vrule#& \hfil#\hfil& 
\vrule#\tabskip=0pt\cr
\tablerule
\omit&height4pt&\omit&&\omit&&\omit&&\omit&\cr
&&\omit\hidewidth \# &&\omit\hidewidth {\rm Vertex Factor}\hidewidth&& 
\omit\hidewidth \#\hidewidth&& \omit\hidewidth {\rm Vertex Factor}
\hidewidth&\cr
\omit&height4pt&\omit&&\omit&&\omit&&\omit&\cr
\tablerule
\omit&height2pt&\omit&&\omit&&\omit&&\omit&\cr
&& 1 && $\frac12 \kappa H \Omega^3 \eta^{\alpha_1 \beta_1}  \eta^{\alpha_2 
\beta_2}  \partial_2^{(\alpha_3}  t^{\beta_3)}$ 
&& 22 && $\frac12 \kappa \Omega^2 \eta^{\alpha_2 (\alpha_3}  \eta^{\beta_3) 
\beta_2}  \partial_3^{(\alpha_1}  \partial_1^{\beta_1)}$ &\cr
\omit&height2pt&\omit&&\omit&&\omit&&\omit&\cr
\tablerule
\omit&height2pt&\omit&&\omit&&\omit&&\omit&\cr
&& 2 && $\frac12 \kappa H \Omega^3 \eta^{\alpha_2 \beta_2}  \eta^{\alpha_3 
\beta_3}  \partial_3^{(\alpha_1}  t^{\beta_1)}$ 
&& 23 && $\frac12 \kappa \Omega^2 \eta^{\alpha_3 (\alpha_1}  \eta^{\beta_1) 
\beta_3}  \partial_1^{(\alpha_2}  \partial_2^{\beta_2)}$ &\cr
\omit&height2pt&\omit&&\omit&&\omit&&\omit&\cr
\tablerule
\omit&height2pt&\omit&&\omit&&\omit&&\omit&\cr
&& 3 && $\frac12 \kappa H \Omega^3 \eta^{\alpha_3 \beta_3}  \eta^{\alpha_1 
\beta_1}  \partial_1^{(\alpha_2}  t^{\beta_2)}$ 
&& 24 && $\frac12 \kappa \Omega^2 \partial_2^{(\alpha_1}  \eta^{\beta_1) 
(\alpha_3}  \partial_3^{\beta_3)}  \eta^{\alpha_2 \beta_2}$ &\cr
\omit&height2pt&\omit&&\omit&&\omit&&\omit&\cr
\tablerule
\omit&height2pt&\omit&&\omit&&\omit&&\omit&\cr
&& 4 && $-\kappa H \Omega^3 \eta^{\alpha_1 (\alpha_2}  \eta^{\beta_2) \beta_1}
 \partial_2^{(\alpha_3}  t^{\beta_3)}$ 
&& 25 && $\frac12 \kappa \Omega^2 \partial_3^{(\alpha_2}  \eta^{\beta_2) 
(\alpha_1}  \partial_1^{\beta_1)}  \eta^{\alpha_3 \beta_3}$ &\cr
\omit&height2pt&\omit&&\omit&&\omit&&\omit&\cr
\tablerule
\omit&height2pt&\omit&&\omit&&\omit&&\omit&\cr
&& 5 && $-\kappa H \Omega^3 \eta^{\alpha_2 (\alpha_3}  \eta^{\beta_3) \beta_2}
 \partial_3^{(\alpha_1}  t^{\beta_1)}$ 
&& 26 && $\frac12 \kappa \Omega^2 \partial_1^{(\alpha_3}  \eta^{\beta_3) 
(\alpha_2}  \partial_2^{\beta_2)}  \eta^{\alpha_1 \beta_1}$ &\cr
\omit&height2pt&\omit&&\omit&&\omit&&\omit&\cr
\tablerule
\omit&height2pt&\omit&&\omit&&\omit&&\omit&\cr
&& 6 && $-\kappa H \Omega^3 \eta^{\alpha_3 (\alpha_1}  \eta^{\beta_1) \beta_3}
 \partial_1^{(\alpha_2}  t^{\beta_2)}$ 
&& 27 && $\frac12 \kappa \Omega^2 \partial_2^{(\alpha_1}  \eta^{\beta_1) 
(\alpha_2}  \partial_3^{\beta_2)}  \eta^{\alpha_3 \beta_3}$ &\cr
\omit&height2pt&\omit&&\omit&&\omit&&\omit&\cr
\tablerule
\omit&height2pt&\omit&&\omit&&\omit&&\omit&\cr
&& 7 && $-\kappa H \Omega^3 t^{(\alpha_3}  \eta^{\beta_3) (\alpha_1}
 \partial_2^{\beta_1)}  \eta^{\alpha_2 \beta_2}$ 
&& 28 && $\frac12 \kappa \Omega^2 \partial_3^{(\alpha_2}  \eta^{\beta_2) 
(\alpha_3}  \partial_1^{\beta_3)}  \eta^{\alpha_1 \beta_1}$ &\cr
\omit&height2pt&\omit&&\omit&&\omit&&\omit&\cr
\tablerule
\omit&height2pt&\omit&&\omit&&\omit&&\omit&\cr
&& 8 && $-\kappa H \Omega^3 t^{(\alpha_1}  \eta^{\beta_1) (\alpha_2} 
 \partial_3^{\beta_2)}  \eta^{\alpha_3 \beta_3}$ 
&& 29 && $\frac12 \kappa \Omega^2 \partial_1^{(\alpha_3}  \eta^{\beta_3) 
(\alpha_1}  \partial_2^{\beta_1)}  \eta^{\alpha_2 \beta_2}$ &\cr
\omit&height2pt&\omit&&\omit&&\omit&&\omit&\cr
\tablerule
\omit&height2pt&\omit&&\omit&&\omit&&\omit&\cr
&& 9 && $-\kappa H \Omega^3 t^{(\alpha_2}  \eta^{\beta_2) (\alpha_3} 
 \partial_1^{\beta_3)}  \eta^{\alpha_1 \beta_1}$ 
&& 30 && $\frac18 \kappa \Omega^2 \eta^{\alpha_1 \beta_1}  \eta^{\alpha_2 
\beta_2}  \eta^{\alpha_3 \beta_3}  \partial_2 \cdot \partial_3$ &\cr
\omit&height2pt&\omit&&\omit&&\omit&&\omit&\cr
\tablerule
\omit&height2pt&\omit&&\omit&&\omit&&\omit&\cr
&& 10 && $\frac14 \kappa \Omega^2 \eta^{\alpha_1 \beta_1}  \partial_3^{
(\alpha_2}  \eta^{\beta_2) (\alpha_3}  \partial_2^{\beta_3)}$ 
&& 31 && $\frac14 \kappa \Omega^2 \eta^{\alpha_1 \beta_1}  \eta^{\alpha_2 
\beta_2}  \eta^{\alpha_3 \beta_3}  \partial_3 \cdot \partial_1$ &\cr
\omit&height2pt&\omit&&\omit&&\omit&&\omit&\cr
\tablerule
\omit&height2pt&\omit&&\omit&&\omit&&\omit&\cr
&& 11 && $\frac14 \kappa \Omega^2 \eta^{\alpha_2 \beta_2}  \partial_1^{
(\alpha_3}  \eta^{\beta_3) (\alpha_1}  \partial_3^{\beta_1)}$ 
&& 32 && $-\frac12 \kappa \Omega^2 \eta^{\alpha_1 (\alpha_2}  \eta^{\beta_2) 
\beta_1}  \eta^{\alpha_3 \beta_3}  \partial_2 \cdot \partial_3$ &\cr
\omit&height2pt&\omit&&\omit&&\omit&&\omit&\cr
\tablerule
\omit&height2pt&\omit&&\omit&&\omit&&\omit&\cr
&& 12 && $\frac14 \kappa \Omega^2 \eta^{\alpha_3 \beta_3}  \partial_2^{
(\alpha_1}  \eta^{\beta_1) (\alpha_2}  \partial_1^{\beta_2)}$ 
&& 33 && $-\frac12 \kappa \Omega^2 \eta^{\alpha_2 (\alpha_3}  \eta^{\beta_3) 
\beta_2}  \eta^{\alpha_1 \beta_1}  \partial_3 \cdot \partial_1$ &\cr
\omit&height2pt&\omit&&\omit&&\omit&&\omit&\cr
\tablerule
\omit&height2pt&\omit&&\omit&&\omit&&\omit&\cr
&& 13 && $-\kappa \Omega^2 \partial_3^{(\alpha_1}  \eta^{\beta_1) (\alpha_2} 
 \eta^{\beta_2) (\alpha_3}  \partial_2^{\beta_3)}$ 
&& 34 && $-\frac12 \kappa \Omega^2 \eta^{\alpha_3 (\alpha_1}  \eta^{\beta_1) 
\beta_3}  \eta^{\alpha_2 \beta_2}  \partial_1 \cdot \partial_2$ &\cr
\omit&height2pt&\omit&&\omit&&\omit&&\omit&\cr
\tablerule
\omit&height2pt&\omit&&\omit&&\omit&&\omit&\cr
&& 14 && $-\kappa \Omega^2 \partial_1^{(\alpha_2}  \eta^{\beta_2) (\alpha_3} 
 \eta^{\beta_3) (\alpha_1}  \partial_3^{\beta_1)}$ 
&& 35 && $-\frac14 \kappa \Omega^2 \partial_2^{(\alpha_1}  \partial_3^{
\beta_1)}  \eta^{\alpha_2 \beta_2}  \eta^{\alpha_3 \beta_3}$ &\cr
\omit&height2pt&\omit&&\omit&&\omit&&\omit&\cr
\tablerule
\omit&height2pt&\omit&&\omit&&\omit&&\omit&\cr
&& 15 && $-\kappa \Omega^2 \partial_2^{(\alpha_3}  \eta^{\beta_3) (\alpha_1} 
 \eta^{\beta_1) (\alpha_2}  \partial_1^{\beta_2)}$ 
&& 36 && $-\frac12 \kappa \Omega^2 \partial_3^{(\alpha_2}  \partial_1^{
\beta_2)}  \eta^{\alpha_3 \beta_3}  \eta^{\alpha_1 \beta_1}$ &\cr
\omit&height2pt&\omit&&\omit&&\omit&&\omit&\cr
\tablerule
\omit&height2pt&\omit&&\omit&&\omit&&\omit&\cr
&& 16 && $-\frac12 \kappa \Omega^2 \partial_3^{(\alpha_2}  \eta^{\beta_2) 
(\alpha_1}  \eta^{\beta_1) (\alpha_3}  \partial_2^{\beta_3)}$ 
&& 37 && $-\frac18 \kappa \Omega^2 \eta^{\alpha_1 \beta_1}  \eta^{\alpha_2 
(\alpha_3}  \eta^{\beta_3) \beta_2}  \partial_2 \cdot \partial_3$ &\cr
\omit&height2pt&\omit&&\omit&&\omit&&\omit&\cr
\tablerule
\omit&height2pt&\omit&&\omit&&\omit&&\omit&\cr
&& 17 && $-\kappa \Omega^2 \partial_1^{(\alpha_3}  \eta^{\beta_3) (\alpha_2} 
 \eta^{\beta_2) (\alpha_1}  \partial_3^{\beta_1)}$ 
&& 38 && $-\frac14 \kappa \Omega^2 \eta^{\alpha_2 \beta_2}  \eta^{\alpha_3 
(\alpha_1}  \eta^{\beta_1) \beta_3}  \partial_3 \cdot \partial_1$ &\cr
\omit&height2pt&\omit&&\omit&&\omit&&\omit&\cr
\tablerule
\omit&height2pt&\omit&&\omit&&\omit&&\omit&\cr
&& 18 && $-\frac14 \kappa \Omega^2 \eta^{\alpha_1 \beta_1}  \eta^{\alpha_2 
\beta_2}  \partial_2^{(\alpha_3}  \partial_3^{\beta_3)}$ 
&& 39 && $\frac12 \kappa \Omega^2 \eta^{\alpha_1) (\alpha_2}  \eta^{\beta_2) 
(\alpha_3}  \eta^{\beta_3) (\beta_1}  \partial_2 \cdot \partial_3$ &\cr
\omit&height2pt&\omit&&\omit&&\omit&&\omit&\cr
\tablerule
\omit&height2pt&\omit&&\omit&&\omit&&\omit&\cr
&& 19 && $-\frac14 \kappa \Omega^2 \eta^{\alpha_2 \beta_2}  \eta^{\alpha_3 
\beta_3}  \partial_3^{(\alpha_1}  \partial_1^{\beta_1)}$ 
&& 40 && $\kappa \Omega^2 \eta^{\alpha_1) (\alpha_2}  \eta^{\beta_2) 
(\alpha_3}  \eta^{\beta_3) (\beta_1}  \partial_3 \cdot \partial_1$ &\cr
\omit&height2pt&\omit&&\omit&&\omit&&\omit&\cr
\tablerule
\omit&height2pt&\omit&&\omit&&\omit&&\omit&\cr
&& 20 && $-\frac14 \kappa \Omega^2 \eta^{\alpha_3 \beta_3}  \eta^{\alpha_1 
\beta_1}  \partial_1^{(\alpha_2}  \partial_2^{\beta_2)}$ 
&& 41 && $\frac14 \kappa \Omega^2 \partial_2^{(\alpha_1}  \partial_3^{
\beta_1)}  \eta^{\alpha_2 (\alpha_3}  \eta^{\beta_3) \beta_2}$ &\cr
\omit&height2pt&\omit&&\omit&&\omit&&\omit&\cr
\tablerule
\omit&height2pt&\omit&&\omit&&\omit&&\omit&\cr
&& 21 && $\frac12 \kappa \Omega^2 \eta^{\alpha_1 (\alpha_2}  \eta^{\beta_2) 
\beta_1}  \partial_2^{(\alpha_3}  \partial_3^{\beta_3)}$
&& 42 && $\frac12 \kappa \Omega ^2 \partial_3^{(\alpha_2}  \partial_1^{
\beta_2)}  \eta^{\alpha_3 (\alpha_1}  \eta^{\beta_1) \beta_3}$ &\cr 
\omit&height2pt&\omit&&\omit&&\omit&&\omit&\cr
\tablerule}}

\caption{Vertex factors contracted into $\psi_{\alpha_1\beta_1} \psi_{\alpha_2
\beta_2} \psi_{\alpha_3\beta_3}$ with $\psi_{\alpha_1\beta_1}$ external.}

\end{table}

When the external leg attaches to the pseudo-graviton field the interaction
makes contributions to $\alpha(t)$ and $\gamma(t)$. Because $\psi = -\psi_{00} 
+ \psi_{ii}$ these have opposite signs. The contribution for $\alpha(t)$ is:
\begin{eqnarray}
\lefteqn{\left\{i \cdot \kappa \cdot -\frac{i}4 \kappa m^2 \Omega^4 \cdot
\langle 0 \vert T\left[ \phi(x) \phi(x')\right] \vert 0 \rangle\right\}_{x'=x}
= \frac14 \kappa^2 m^2 \Omega^4 i \Delta_{\gamma}(x;x) \; ,} \qquad \qquad \\
& & \qquad \stackrel{IR}{\longrightarrow} -\frac34 \kappa^2 \dot{H} 
a^4_0(t) \cdot {-\dot{H} \over H^2(t)} \cdot {1 \over 24 \pi^2} \left({H_I^6 - 
H^6(t) \over \dot{H}^2}\right) \; , \qquad \\
& & \qquad = {\kappa^2 \over 32 \pi^2} \left({H_I^6 - H^6(t) \over 
H^2(t)}\right) a_0^4(t) \; .
\end{eqnarray}
In the notation used at the end of Section 5 there are contributions for $N=-2$
and $N=+4$ with coefficients of $\kappa^2/(32\pi^2)$ times $+H^6_I$ and $-1$,
respectively.

Tables 1-3 give the various cubic interaction vertices. Note that those of 
Table 1 have been partially symmetrized by making the external leg attach to
pseudo-graviton \#1. It should also be noted that derivatives with respect to
the attached field are interpreted through integration by parts (IBP) as acting
on the entire result. Further, spatial translation invariance (STI) allows the
result to depend only upon the conformal time after the expectation value is 
taken. For example, in vertex \# 3 of Table 1, the derivative with respect to 
line \#1 is interpreted as follows:
\begin{equation}
\partial_1^{\alpha_2} \stackrel{IBP}{\longrightarrow} - {\partial \over 
\partial x_{\alpha_2}} \stackrel{STI}{\longrightarrow} t^{\alpha_2} {d \over
d\eta} \; .
\end{equation}
Because of the $\eta^{\alpha_1\beta_1}$ vertex \#3 contributes to $\alpha(t)$
and $\gamma(t)$ with opposite signs. The contribution to $\alpha(t)$ is:
\begin{eqnarray}
\lefteqn{{d \over d\eta} \left\{i \cdot \kappa \cdot \frac{i}2 \kappa H 
\Omega^3 \cdot \langle 0 \vert T\left[ \psi_{\alpha_2\beta_2}(x) \psi_{\alpha_3
\beta_3}(x')\right] \vert 0 \rangle t^{\alpha_2} t^{\beta_2} \eta^{\alpha_3
\beta_3} \right\}_{x' = x}} 
\nonumber \\
& & \qquad \qquad = {d \over d\eta} \left\{-\frac12 \kappa^2 H \Omega^3 \cdot
2 i \Delta_{\alpha}(x;x')\right\}_{x'=x} \; ,\\
& & \qquad \qquad \stackrel{IR}{\longrightarrow} -3 \kappa^2 H^2(t) a^4_0(t) 
\cdot {\dot{H}^2 \over H^4(t)} \cdot {1 \over 24 \pi^2} \left({H_I^6 - H^6(t) 
\over \dot{H}^2}\right) \; , \\
& & \qquad \qquad = - {\kappa^2 \over 8 \pi^2} \left({H_I^6 - H^6(t) \over 
H^2(t)}\right) a_0^4(t) \; .
\end{eqnarray}
 
\begin{table}

\vbox{\tabskip=0pt \offinterlineskip
\def\tablerule{\noalign{\hrule}}
\halign to450pt {\strut#& \vrule#\tabskip=1em plus2em& 
\hfil#& \vrule#& \hfil#\hfil& \vrule#& \hfil#& \vrule#& \hfil#\hfil& 
\vrule#\tabskip=0pt\cr
\tablerule
\omit&height4pt&\omit&&\omit&&\omit&&\omit&\cr
&&\omit\hidewidth \# 
&&\omit\hidewidth {\rm Vertex Factor}\hidewidth&& 
\omit\hidewidth \#\hidewidth&& 
\omit\hidewidth {\rm Vertex Factor}
\hidewidth&\cr
\omit&height4pt&\omit&&\omit&&\omit&&\omit&\cr
\tablerule
\omit&height2pt&\omit&&\omit&&\omit&&\omit&\cr
&& 1 && $- \kappa \Omega^2 \eta^{\alpha_2 (\alpha_1} \eta^{\beta_1) 
\alpha_3} \partial_2 \cdot \partial_3$ 
&& 6 && $\frac12 \kappa \Omega^2 \eta^{\alpha_1 \beta_1} \partial_2^{
\alpha_2} \partial_1^{\alpha_3}$ &\cr
\omit&height2pt&\omit&&\omit&&\omit&&\omit&\cr
\tablerule
\omit&height2pt&\omit&&\omit&&\omit&&\omit&\cr
&& 2 && $- \kappa \Omega^2 \eta^{\alpha_3 (\alpha_1} \partial_2^{\beta_1)} 
\partial_3^{\alpha_2}$ 
&& 7 && $-\kappa H \Omega^3 \eta^{\alpha_1 \beta_1} \partial_2^{\alpha_2} 
t^{\alpha_3}$ &\cr
\omit&height2pt&\omit&&\omit&&\omit&&\omit&\cr
\tablerule
\omit&height2pt&\omit&&\omit&&\omit&&\omit&\cr
&& 3 && $- \kappa \Omega^2 \eta^{\alpha_2 (\alpha_1} \partial_2^{\beta_1)} 
\partial_1^{\alpha_3}$ 
&& 8 && $-2 \kappa H \Omega^3 \eta^{\alpha_3 (\alpha_1} \partial_3^{
\beta_1)} t^{\alpha_2}$ &\cr
\omit&height2pt&\omit&&\omit&&\omit&&\omit&\cr
\tablerule
\omit&height2pt&\omit&&\omit&&\omit&&\omit&\cr
&& 4 && $2 \kappa H \Omega^3 \eta^{\alpha_2 (\alpha_1} \partial_2^{
\beta_1)} t^{\alpha_3}$ 
&& 9 && $-\kappa H \Omega^3 \eta^{\alpha_1 \beta_1} \partial_1^{\alpha_3} 
t^{\alpha_2}$ &\cr
\omit&height2pt&\omit&&\omit&&\omit&&\omit&\cr
\tablerule
\omit&height2pt&\omit&&\omit&&\omit&&\omit&\cr
&& 5 && $\kappa \Omega^2 \eta^{\alpha_3 (\alpha_1} \partial_3^{\beta_1)} 
\partial_2^{\alpha_2}$ 
&& 10 && $2 \kappa H^2 \Omega^4 \eta^{\alpha_1 \beta_1} t^{\alpha_2} 
t^{\alpha_3}$ &\cr
\omit&height2pt&\omit&&\omit&&\omit&&\omit&\cr
\tablerule}}

\caption{Vertex factors contracted into $\psi_{\alpha_1\beta_1} 
{\overline \omega}_{\alpha_2} \omega_{\alpha_3}$.}

\end{table}

\begin{table}

\vbox{\tabskip=0pt \offinterlineskip
\def\tablerule{\noalign{\hrule}}
\halign to450pt {\strut#& \vrule#\tabskip=1em plus2em& 
\hfil#& \vrule#& \hfil#\hfil& \vrule#& \hfil#& \vrule#& \hfil#\hfil& 
\vrule#\tabskip=0pt\cr
\tablerule
\omit&height4pt&\omit&&\omit&&\omit&&\omit&\cr
&&\omit\hidewidth \# 
&&\omit\hidewidth {\rm Interaction}\hidewidth&& 
\omit\hidewidth \#\hidewidth&& 
\omit\hidewidth {\rm Interaction}
\hidewidth&\cr
\omit&height4pt&\omit&&\omit&&\omit&&\omit&\cr
\tablerule
\omit&height2pt&\omit&&\omit&&\omit&&\omit&\cr
&& 1 && $\frac18 \kappa^2 \varphi_0^{\prime} \Omega^2 \phi^{\prime} \psi^2$
&& 6 && $\frac12 \kappa \Omega^2 \phi_{,\rho} \phi_{,\sigma} 
\psi^{\rho\sigma}$ &\cr 
\omit&height2pt&\omit&&\omit&&\omit&&\omit&\cr
\tablerule
\omit&height2pt&\omit&&\omit&&\omit&&\omit&\cr
&& 2 && $-\frac14 \kappa^2 \varphi_0^{\prime} \Omega^2 \phi^{\prime} 
\psi^{\rho\sigma} \psi_{\rho\sigma}$
&& 7 && $-\frac18 \kappa^2 m^2 \varphi_0 \Omega^4 \phi \psi^2$ &\cr
\omit&height2pt&\omit&&\omit&&\omit&&\omit&\cr
\tablerule
\omit&height2pt&\omit&&\omit&&\omit&&\omit&\cr
&& 3 && $-\frac12 \kappa^2 \varphi_0^{\prime} \Omega^2 \phi_{,\rho} 
\psi^{\rho}_{~0} \psi$
&& 8 && $\frac14 \kappa^2 m^2 \varphi_0 \Omega^4 \phi \psi^{\rho\sigma}
\psi_{\rho\sigma}$ &\cr 
\omit&height2pt&\omit&&\omit&&\omit&&\omit&\cr
\tablerule
\omit&height2pt&\omit&&\omit&&\omit&&\omit&\cr
&& 4 && $\kappa^2 \varphi_0^{\prime} \Omega^2 \phi_{,\rho} \psi^{\rho\sigma}
\psi_{\sigma 0}$
&& 9 && $-\frac14 \kappa m^2 \Omega^4 \phi^2 \psi$ &\cr 
\omit&height2pt&\omit&&\omit&&\omit&&\omit&\cr
\tablerule
\omit&height2pt&\omit&&\omit&&\omit&&\omit&\cr
&& 5 && $-\frac14 \kappa \Omega^2 \phi_{,\rho} \phi^{,\rho} \psi$
&& 10 && $\kappa^2 \varphi_0^{\prime} \Omega^2 \phi_{,\rho} {\overline 
\omega}_0 \omega^{\rho}$ &\cr 
\omit&height2pt&\omit&&\omit&&\omit&&\omit&\cr
\tablerule}}

\caption{Cubic interactions involving $\phi$.}

\end{table}

When the contributions from all vertices are summed the amputated 1-point 
functions are:
\begin{eqnarray}
\alpha(t) & \stackrel{IR}{\longrightarrow} & - {3 \kappa^2 \over 32 \pi^2}
\left({H_I^6 - H^6(t) \over H^2(t)}\right) a_0^4(t) \; ,\\
\gamma(t) & \stackrel{IR}{\longrightarrow} & + {3 \kappa^2 \over 32 \pi^2}
\left({H_I^6 - H^6(t) \over H^2(t)}\right) a_0^4(t) \; ,\\
\delta(t) & \stackrel{IR}{\longrightarrow} & - {\kappa^2 \over 8 \pi^2}
\left({H_I^6 - H^6(t) \over H^2(t)}\right) \left({\sqrt{-\dot{H}} \over H(t)}
\right) a_0^4(t) \; .
\end{eqnarray}
In the notation at the end of Section 5 this corresponds to the following
coefficients:
\begin{eqnarray}
\alpha_{-2} = - \gamma_{-2} = \frac34 \delta_{-2} & = & - {3 \kappa^2 \over
32 \pi^2} H_I^6 \; ,\\
\alpha_{+4} = - \gamma_{+4} = \frac34 \delta_{+4} & = & {3 \kappa^2 \over
32 \pi^2} \; .
\end{eqnarray}
From (\ref{eq:HeffN}-\ref{eq:PhiN}) we see that the cosmological expansion 
rate and the quantum-corrected scalar are:
\begin{eqnarray}
H_{\rm eff}({\overline t}) & = & H({\overline t}) \left\{1 - {\kappa^2 \over 
72 \pi^2} \left({H_I^6 - 3 H_I^3 H^3({\overline t}) + 2 H^6({\overline t})
\over H^4({\overline t})}\right) + \dots \right\} \; , \quad \\
\Phi({\overline t}) & = & \varphi_0({\overline t}) \left\{1 + {\kappa^2 \over 
576 \pi^2} \left({H_I^6 - 2 H_I^3 H^3({\overline t}) + H^6({\overline t})
\over H^4({\overline t})}\right) + \dots \right\} \; . \quad
\end{eqnarray}

\section{Summary and discussion}

This paper is first of all a check on the calculation of Mukhanov, Abramo and
Brandenberger. We certainly agree with the sign and the leading time dependence
that can be inferred for $H_{\rm eff}({\overline t})$ and $\Phi({\overline t})$
from their published results \cite{Mukh,Abra}. From unpublished work we see 
that even the numerical factors agree. It is worth emphasizing that we employed
the standard formalism of covariant quantization while they used a truncated 
version of the canonical formalism. The two calculations were also done in 
completely different gauges: we added a covariant gauge fixing term whereas 
they used a physical gauge. It would be difficult to imagine two more 
completely different calculational schemes. Yet we got the same results in the 
end, for both $H_{\rm eff}({\overline t})$ and $\Phi({\overline t})$. {\bf This
is an enormously powerful check on the validity of their work and on the 
physical reality of the effect.} If this back-reaction is a gauge chimera it is
a remarkably consistent one.

Our motivation for this work was the excellent questions posed earlier this 
year by Unruh \cite{Unru}. Although our calculation is itself a sort of answer 
we have analyzed some of his arguments more generally in Section 3. In 
particular, it turns out that the seeming disagreement between Unruh's long 
wavelength solutions and those of Mukhanov {\it et al.} derives from different
definitions for what is ``zeroth order'' when expanding in powers of the wave
number. If one keeps higher terms in $k^2$ it turns out that Unruh's second
solution implies an order $k^2$ result for the variable used by Mukhanov {\it
et al.}, and this result is just $4 k^2/\kappa^2$ times their second solution.
So neither of the long wavelength solutions of Mukhanov {\it et al.} is 
unphysical.

It is important to understand that a compelling physical mechanism underlies
the back-reaction of Mukhanov {\it et al.}. It is the self-gravitation between 
superadiabatically amplified long wavelength modes. A simple physical model is 
that virtual particles whose wavelengths are comparable to the Hubble radius 
become trapped in the expansion of spacetime and are not able to recombine. As 
the particles are pulled apart their long range gravitational potentials fill 
the intervening space, adding with the potentials of earlier pairs. Because 
gravitation is attractive these potentials resist the further expansion of 
spacetime, thereby slowing inflation. There is absolutely no question that this
process should occur for quanta, such as gravitons and minimally coupled 
scalars, which lack conformal invariance but are still massless on the Hubble 
scale. The only issues concern the strength of the back-reaction, its time 
dependence, and whether or not it can eventually stop inflation.

The analogous back-reaction has already been demonstrated for gravitons when 
inflation is driven by a positive bare cosmological constant \cite{Tsam}. In
this case it comes at two loops because the pair creation event is already one
loop and the absence of linearized mixing between the dynamical spin 2 
gravitons and the spin 0 gravitational potentials postpones self-gravitation to
next order. When the superadiabatically amplified quanta are themselves scalar
their mixing with the spin 0 gravitational potentials allows self-gravitation 
to occur at one loop order. However, there does not have to be such a one loop
effect \cite{AbWo}. The feature which seems to distinguish those scalar-driven 
models which show slowing at one loop from those which slow at two loops is the
rate at which superadiabatic amplification injects 0-point energy. If this is
less than or equal to the physical 3-volume's inflation then there is no one
loop effect; if superadiabatic amplification injects 0-point energy faster than
the 3-volume inflates to absorb it then there is a one loop effect.

Of course gravitons presumably drive a two loop effect in this model as well.
There may also be significant scalar effects at higher loops. Higher loop 
processes are interesting in that they derive from the coherent superposition 
of interactions over the invariant volume of the past lightcone, which can grow
arbitrarily large. There is no barrier to considering such questions in the 
covariant formalism we have developed. The formalism of Mukhanov {\it et al.} 
would have to be extended to make this possible.

It may be of general interest that we were able to extend the Feynman rules of 
Iliopoulos {\it et al.} \cite{Ilio} so that they apply to an arbitrary scalar 
potential. This was the work of Section 4. Of course we can only express the 
propagators as mode sums, where even the mode functions remain to be 
determined. But we have shown in Section 6 how these mode functions can be
usefully expanded, both in the ultraviolet and in the infrared. And the 
calculation is an explicit example of how interesting effects can be obtained.
It should now be possible to re-do the two loop computation of Tsamis and 
Woodard \cite{Tsam} for an arbitrary background. This should completely 
determine the effective field equations needed to evolve past the end of 
inflation to arbitrarily late times \cite{TsWo}.

\vskip 1cm
\centerline{\bf Acknowledgments}

It is a pleasure to acknowledge stimulating and informative conversations with 
R. H. Brandenberger and V. F. Mukhanov. We are also grateful to the University 
of Crete for its hospitality at the inception of this project. This work was 
partially supported by DOE contract DE-FG02-97ER\-41029, by NSF grant 
94092715, by NATO grant CRG-971166 and by the Institute for Fundamental Theory.

\end{document}